\documentclass[11pt,a4paper]{article} 
\pdfoutput=1
\usepackage{jheppub,hyperref,mdwlist}
\usepackage{amsmath}
\usepackage{graphicx}
\usepackage{subfigure}
\usepackage{amssymb}
\usepackage{xcolor}
\usepackage{multirow}
\usepackage{tablefootnote}
\usepackage{cancel}
\usepackage{color}
\usepackage{listings}
\usepackage{xcolor}
\usepackage{float}
\usepackage{slashed}
\usepackage{diagbox}
\usepackage{mathtools}

\usepackage{amsmath}
\usepackage{wrapfig}
\usepackage{soul}

\newcommand{\be}{\begin{equation}}
\newcommand{\ee}{\end{equation}}
\newcommand{\beq}{\begin{equation}}
\newcommand{\eeq}{\end{equation}}
\newcommand{\bea}{\begin{eqnarray}}
\newcommand{\eea}{\end{eqnarray}}
\newcommand{\besp}{\begin{equation}\begin{split}}
\newcommand{\eesp}{\end{split}\end{equation}}

\newcommand{\nn}{\nonumber}

\newcommand{\tr}{\text{tr}}
\newcommand{\Br}{\text{Br}}

\newcommand{\tabincell}[2]{\begin{tabular}{@{}#1@{}}#2\end{tabular}}
\newcommand{\Eq}[1]{Eq.~(\ref{#1})}

\newcommand{\Dfbd}{\mathord{\buildrel{\lower3pt\hbox{$\scriptscriptstyle\leftrightarrow$}}\over {D}_{\mu}}}
\newcommand{\ave}[1]{\left\langle #1\right\rangle}

\def\hc{{\rm h.c.}}

\def\mG{\mathcal{G}}
\def\mH{\mathcal{H}}

\def\mL{\mathcal{L}}

\def\mO{\mathcal{O}}

\def\d{\text{d}}

\def\0{\textbf{0}}
\def\1{\textbf{1}}
\def\2{\textbf{2}}
\def\3{\textbf{3}}
\def\4{\textbf{4}}
\def\5{\textbf{5}}
\def\6{\textbf{6}}
\def\7{\textbf{7}}
\def\8{\textbf{8}}
\def\9{\textbf{9}}

\begin{document}

\title{Composite resonances at a 10 TeV muon collider}

\author[a,b]{Da Liu,}
\author[c,d,e]{Lian-Tao Wang,}
\author[f]{and Ke-Pan Xie}
\affiliation[a]{PITT PACC, University of Pittsburgh, Pittsburgh, PA, USA}
\affiliation[b]{Center for Quantum Mathematics and Physics (QMAP), University of California, Davis, CA 95616, USA}
\affiliation[c]{Enrico Fermi Institute, The University of Chicago, 5640 S Ellis Ave, Chicago, IL 60637, USA}
\affiliation[d]{Department of Physics, The University of Chicago, 5640 S Ellis Ave, Chicago, IL 60637, USA}
\affiliation[e]{Kavli Institute for Cosmological Physics, The University of Chicago, 5640 S Ellis Ave, Chicago, IL 60637, USA}
\affiliation[f]{School of Physics, Beihang University, Beijing 100191, China}

\emailAdd{dal339@pitt.edu}
\emailAdd{liantaow@uchicago.edu}
\emailAdd{kpxie@buaa.edu.cn}

\abstract{We investigate the reach for resonances of the composite Higgs models at a 10 TeV $\mu^+\mu^-$ collider with up to 10 ab$^{-1}$ luminosity. The strong dynamics sector is modeled by the minimal coset $SO(5)/SO(4)$, where vector resonances are in $(\3, \1)$ of $SO(4)$ and fermions are in $(\2, \2)$. Various production and decay channels are studied. For the spin-1 resonances, the projections are made based on the radiative return and vector boson fusion production channels. The muon collider can cover most of the kinematically allowed mass range and can measure the coupling $g_\rho$ to percent level. For the fermionic resonances (i.e. the top partners), pair production easily covers the resonance mass below 5 TeV, while single production extends the reach to 6 TeV for a small $\xi = 0.015$.
}
\maketitle
\flushbottom

\section{Introduction}

Recently, there has been a growing interest in high energy muon colliders~\cite{Accettura:2023ked,MuonCollider:2022nsa,Delahaye:2019omf}, inspired by the physics potential in both precision measurement and resonance searches~\cite{Chiesa:2020awd, Costantini:2020stv, Capdevilla:2020qel, Han:2020uid,Han:2020pif,Han:2020uak,Buttazzo:2020ibd,Yin:2020afe,Buttazzo:2020uzc,Huang:2021nkl,Liu:2021jyc,Capdevilla:2021rwo,Han:2021udl,Capdevilla:2021fmj,Han:2021kes,AlAli:2021let,Asadi:2021gah,Bottaro:2021snn,Qian:2021ihf,Chiesa:2021qpr,Liu:2021akf,Chen:2021pqi,Chen:2022msz,Cesarotti:2022ttv,deBlas:2022aow,Bao:2022onq,Homiller:2022iax,Forslund:2022xjq,Chen:2022yiu,Liu:2023yrb,Forslund:2023reu, Amarkhail:2023xsc,Kwok:2023dck,Li:2023tbx,Yang:2022ilt,Zhang:2023yfg,Yang:2023gos,Zhang:2023ykh,Barger:2023wbg,Cassidy:2023lwd}. Recent encouraging developments include the demonstration of the cooling from MICE collaboration~\cite{MICE:2019jkl}, the establishment of the International Muon Collider Collaboration \cite{Schulte:2021eyr}, and the recent endorsement in the report of US Particle Physics Project Prioritization Panel \cite{p5}. Detailed studies of the physics potential are ongoing. In this paper, we will study the prospects of searching for the heavy resonances in composite Higgs models at a 10 TeV high energy muon collider with an integrated luminosity up to 10 ab$^{-1}$. We will focus on the minimal composite Higgs model (MCHM) that is based on the global symmetry breaking pattern of $SO(5)/SO(4)$ to address the naturalness problem~\cite{Agashe:2004rs,Contino:2006qr}. We  consider the spin-1 resonances of $(\bold 3,\bold 1)$ representation of the unbroken $SO(4)\simeq SU(2)_L\times SU(2)_R$ and fermionic resonances (top partners)  in the $(\bold 2,\bold 2)$ representation. Studying other representations and taking into account the interplay between spin-1 and fermionic resonances is also interesting, which we leave for future work.

This paper is organized as follows. In Section~\ref{sec:spin1}, we will discuss the production and decay channels of the spin-1 resonances, and present the expected reach on the masses and couplings. Then we turn to the top partners in Section~\ref{sec:spin12}, studying the phenomenology and presenting the projected reach. The details of the model under consideration are described in Appendix~\ref{app:model}.

\section{The spin-1 resonances $\rho^{\pm,0} \in (\bold 3,\bold 1)$}
\label{sec:spin1}

In this section, we investigate the potential reach in mass scale and coupling for the $(\bold 3,\bold 1)$ spin-1 $\rho$-resonances of the strong dynamics sector. For a comprehensive description of the models, we refer to Refs.~\cite{Pappadopulo:2014qza,Greco:2014aza,Panico:2015jxa,Liu:2018hum}. A summary of the Lagrangian, relevant mass matrices, and couplings can be found in Appendix~\ref{app:model}.

\subsection{Production and decay}

\begin{figure}
\centering
\subfigure{
\includegraphics[scale=1]{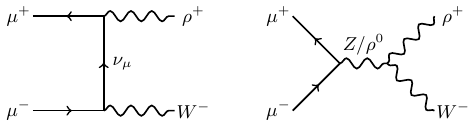}}\qquad\qquad
\subfigure{
\includegraphics[scale=1]{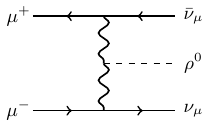}}
\caption{Feynamn diagrams for the single production of $\rho$-resonances  through $\rho V$ associated production (left) and VBF production (right). Other final states are obtained by the proper changes of internal/final particles as required by electric charge conservation. }
\label{fig:rho_production_diagrams}
\end{figure}

The $\rho$ resonances couple to the Standard Model (SM) bosons via the effective Lagrangian:
\be\label{rho_coupling}
\mL\approx \frac{M_\rho^2}{g_\rho}\rho_\mu^a\left(gW_\mu^a-\frac{1}{f^2}H^\dagger\frac{\sigma^a}{2}i\Dfbd H\right),
\ee
where $g_\rho\sim4\pi/\sqrt{N}$ with $N$ the number of ``hyper color'' of the strong dynamics sector. The first term in \Eq{rho_coupling} implies a mixing angle $\theta\approx g/g_\rho$ between $\rho^a$ and $W^a$, and hence $\rho$ couples to the SM fermions with a strength of $\approx g^2/g_\rho$; and the second term provides the $\rho VV$ ($V=W^\pm$, $Z$) vertex through the Goldstone equivalence theorem. As a result, the $\rho^{\pm,0}$ resonances can be singly produced at a $\mu^+\mu^-$ collider mainly via two types of processes: the electroweak (EW) gauge boson associated production (denoted as $\rho V$ production)
\be
\mu^+\mu^-\to\rho^\pm W^\mp,\quad\mu^+\mu^-\to\rho^0Z,~\rho^0\gamma;
\ee
and vector boson fusion (VBF), i.e.
\be\begin{dcases}
~\mu^+\mu^-\to\rho^+\mu^-\bar\nu_\mu,~\rho^-\mu^+\nu_\mu,\quad &(W^\pm Z~{\rm fusion});\\
~\mu^+\mu^-\to\rho^-\nu_\mu\bar\nu_\mu,~\rho^0\mu^+\mu^-,\quad &(W^+W^-/ZZ~{\rm fusion}).
\end{dcases}\ee
The examples of Feynman diagrams of these processes are shown Fig.~\ref{fig:rho_production_diagrams}. The former production mechanism was called ``radiative return" in the literature~\cite{Chakrabarty:2014pja}. 
Its cross section is proportional to the square of the $\rho f\bar f$, thus scales as $g^4/g_\rho^2$ at the leading order (LO). The rate of VBF production can be estimated by using the effective $W$ approximation~\cite{Dawson:1984gx,Kunszt:1987tk,Borel:2012by}, which states that the total cross section can be written as the SM gauge boson PDFs convoluted with the partonic cross sections. We expect that the cross section is dominated by the longitudinal $W^\pm$, $Z$ gauge boson subprocesses as the $\rho$ resonances couple strongly to the longitudinal components. The ratio between the VBF production cross section and the associated production $\rho V$ cross section will scale like $g_\rho^4$ at fixed mass $M_\rho$.

\begin{figure}[t]
\centering
\subfigure{
\includegraphics[scale=1]{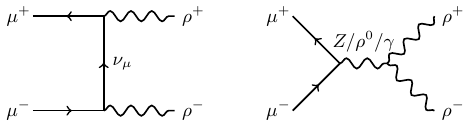}}\qquad\qquad
\subfigure{
\includegraphics[scale=1]{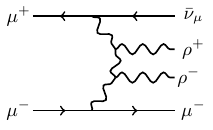}}
\caption{Illustrations for the $\rho$ resonances pair production Feynman diagrams. Left panel: $\mu^+\mu^-$ annihilation; right panel: VBF production. Other final states are obtained by the proper changes of labels of internal/final particles as required by electric charge conservation.}
\label{fig:rho_pair_diagrams}
\end{figure}

In addition to the single production, the $\rho$ resonances can also be produced in pairs at the muon collider. They can be produced either by $\mu^+ \mu^-$ annihilation:
\beq
\mu^+ \mu^- \rightarrow \rho^+ \rho^-, \qquad \mu^+\mu^- \rightarrow \rho^0 \rho^0,
\eeq
or by VBF:
\beq
\mu^+ \mu^- \rightarrow  \rho^+ \rho^- \mu^+ \mu^-,~\rho^+\rho^-\nu_\mu \bar\nu_\mu, \qquad \mu^+ \mu^- \rightarrow  \rho^0 \rho^0 \nu_\mu \bar\nu_\mu.
\eeq
The relevant Feynman diagrams are shown in Fig.~\ref{fig:rho_pair_diagrams}. Among the annihilation processes, the cross section of $\rho^+ \rho^+$ is mainly determined by their EW couplings to the $Z$ and $\gamma$ gauge bosons, which are insensitive to the value of $g_\rho$, especially at large $g_\rho$. On the contrary, the cross section of $\rho^0 \rho^0$ is suppressed by their couplings to the muon and neutrino as $g^8/g_\rho^4$ and is negligible in most of the parameter space. For the $W^+ W^-/ZZ$ fusion processes, we expect that their cross sections are mainly determined by their strong coupling to the longitudinal components of the $W^\pm/Z$-bosons and thus scale like $g_\rho^4$. The $\gamma\gamma$ fusion process is dominated by the electric coupling of the $\rho^\pm$, while the $Z\gamma$ fusion has a $g_\rho^2$ scaling between the two extreme cases.

\begin{figure}[h!]
\centering
\subfigure{
\includegraphics[scale=0.4]{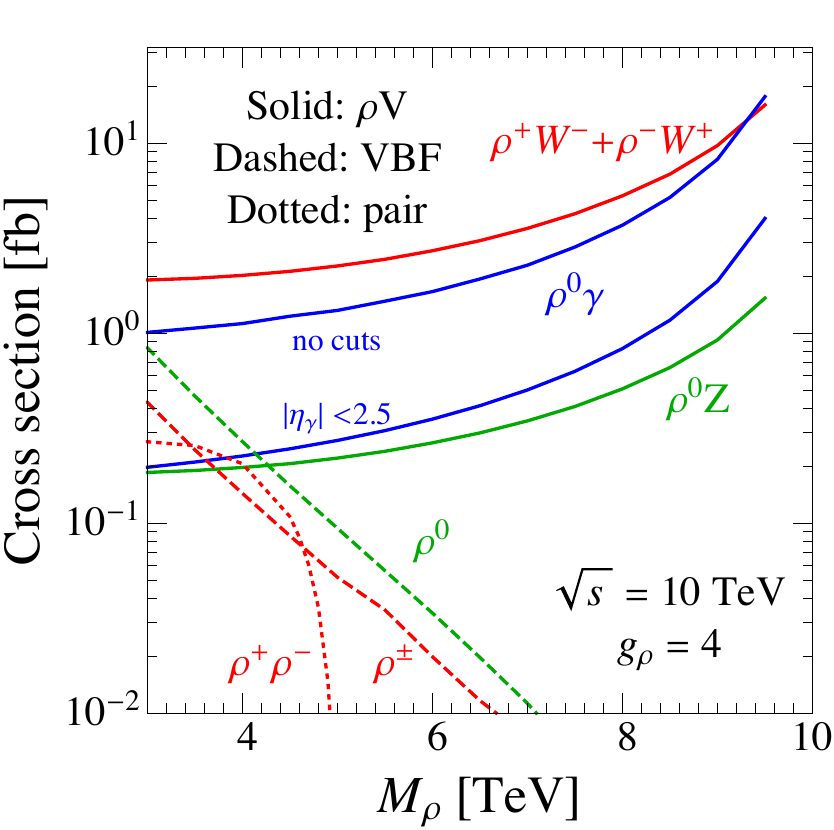}}\qquad\qquad
\subfigure{
\includegraphics[scale=0.4]{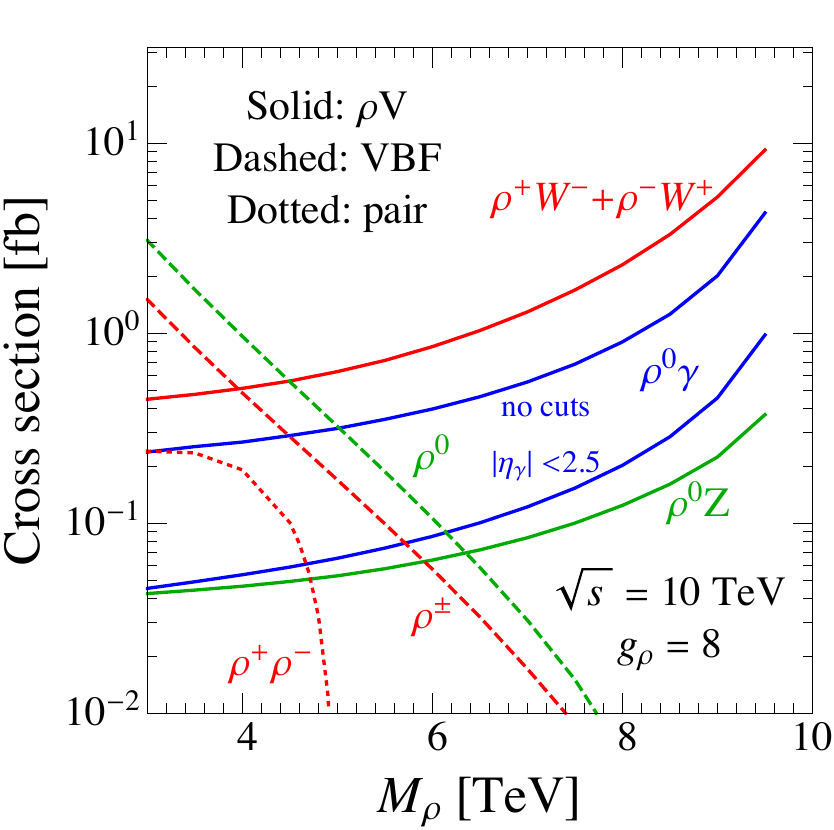}}
\caption{The $\rho$ resonances production rates at muon colliders with $\sqrt{s}=10$ TeV for different $g_\rho$ ( $g_\rho= 4$ for the left panel and $g_\rho = 8$ for the right panel). The   $\xi$ is derived by assuming $a_{\rho}=m_\rho/(g_\rho f)=1/\sqrt{2}$.}
\label{fig:rho_production}
\end{figure}

In Fig.~\ref{fig:rho_production}, we show the production cross sections of the $\rho$ resonances as functions of the mass $M_\rho$ by choosing $g_\rho = 4 ~(8)$ at muon colliders with $\sqrt s=10$ TeV. The cross sections are calculated at LO by using {\tt MadGraph5aMC$@$NLO}~\cite{Alwall:2014hca}. For the $\mu^+\mu^-\to\rho^0\nu_\mu\bar\nu_\mu$ and $\mu^+\mu^-\to\rho^\pm\mu^\mp\nu_\mu$ processes, to select the VBF contribution, we have required the recoil mass to be
\begin{equation}
\label{eq:xscut}
 M_{\rm recoil} = \sqrt{(p_{\mu^+}+p_{\mu^-}-p_\rho)^2} > 1~\text{TeV},
\end{equation}
to eliminate the contributions from $\rho^\pm W^\mp (\to\mu^\pm\nu_\mu)$ or $\rho^0 Z (\to\nu_\mu \bar \nu_\mu)$. For the $\rho^0\gamma$ channel, we have presented two lines:  keeping the finite muon mass $M_\mu\approx105.7$ MeV or putting cuts $|\eta_\gamma|<2.5$ on the photon to remove the collinear divergence.\footnote{The results are quite different, demonstrating a significant contribution from the collinear radiation. A proper treatment is needed to obtain an accurate cross section, which we will leave for future work. Due to its smaller rate, we will not use this channel to set our projection.} For the pair production, due to their smallness, we only show the $\mu^+\mu^-$ annihilation production of $\rho^+ \rho^-$ and will not discuss them further.

We can see that while the VBF production cross section decreases as the mass $M_\rho$ increases, the $\rho V$ associated production cross section increases as the mass $M_\rho$ increases toward the center-of-mass energy of the muons. This behavior can be understood as the infrared (IR) enhancement of $t$-channel processes~\cite{Chakrabarty:2014pja,Han:2021udl}. For example, the $\rho^0\gamma$ associated production rate is
\be\label{IR}
\frac{\d\sigma(\rho^0\gamma)}{\d\cos\theta_\gamma}\approx\frac{\left(g^2e/g_\rho\right)^2}{256\pi^2\sqrt{s}E_\gamma\sin^2\theta_\gamma}\left[\left(1+\frac{M_\rho^2}{s}\right)^2+\left(1-\frac{M_\rho^2}{s}\right)^2\cos^2\theta_\gamma\right],
\ee
where $E_\gamma=(s-M_\rho^2)/(2\sqrt{s})$ and $\theta_\gamma$ are the energy and polar angle of the final state photon, respectively. The cross-section diverges for $E_\gamma\to0 $ and $\sin\theta_\gamma\to0$, corresponding to the soft and collinear IR divergence, respectively. For $\rho^\pm W^\mp$ and $\rho^0Z$ associated productions, the final state SM gauge boson can be treated as massless at a multi-TeV muon collider and hence the cross sections show similar behavior at $s\sim M_\rho^2$. From the above equation, one also obtains the logarithmic enhancement of the production rate at larger $s$ by integrating over the angle $\theta_\gamma$. From the figures, we can infer that at 10 TeV muon collider for $g_\rho = 4~(8)$, the VBF production of neutral $\rho$ resonance dominates over associated production for $M_\rho \lesssim 4.2~(6.4)$ TeV, while for the charged $\rho$ resonance, VBF production dominates for $M_\rho \lesssim 2~(4)$ TeV. 

\begin{figure}
\centering
\subfigure{
\includegraphics[scale=0.4]{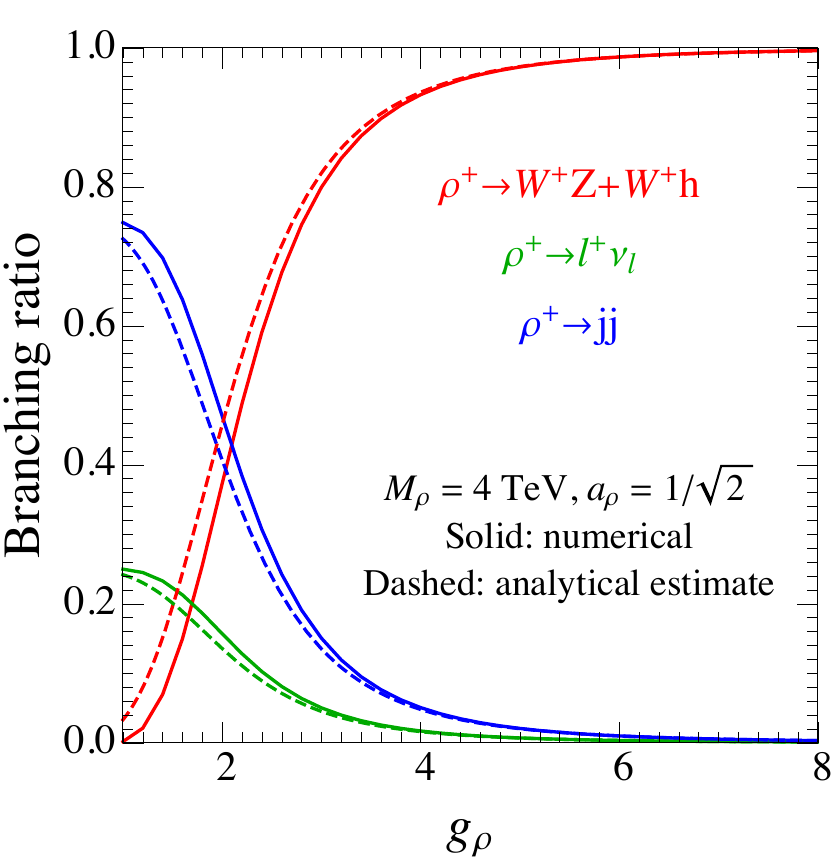}}\qquad\qquad
\subfigure{
\includegraphics[scale=0.4]{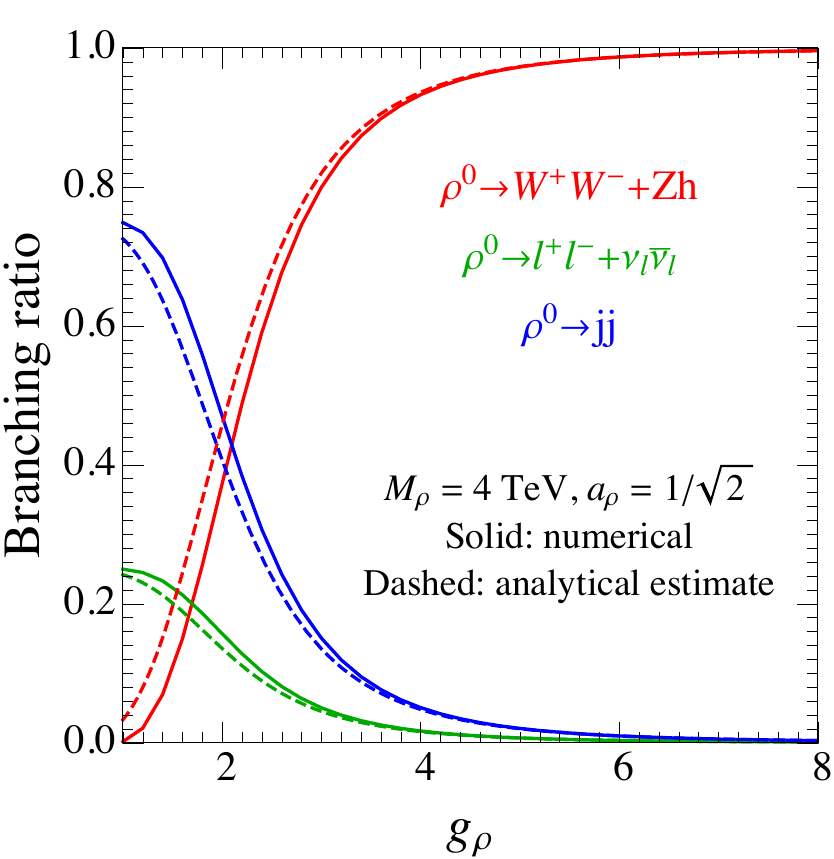}}
\caption{Illustrations for the $\rho$ resonances decay branching ratios without top partners. $M_\rho=4$ TeV and $a_\rho=m_\rho/(g_\rho f)=1/\sqrt{2}$ are adopted as the benchmark.}
\label{fig:rho_decay}
\end{figure}

For the decay of the $\rho$ resonances, we consider the decay channels into SM particles and neglect the possible interactions between the $\rho$ resonances and the top partners. The relevant final states are di-boson ($W^\pm Z/W^\pm h$, $W^+W^-/Zh$), di-leptons ($\ell^-\bar\nu_\ell/\ell^+\nu_\ell$, $\ell^+\ell^-/\nu_\ell\bar\nu_\ell$) and di-quarks ($q\bar q^{(\prime)}$). Neglecting the partial compositeness of the third generation quarks, the SM fermions couple to $\rho$ only via the gauge mixing, yielding a universal coupling $g^2/g_\rho$. On the other hand, the  SM longitudinal gauge bosons and Higgs boson couple to $\rho$ via the Goldstone current, which is proportional to $g_\rho$. As a result, the decay branching ratios of $\rho$ have very clear scaling features, e.g.
\be\begin{split}
\Gamma(\rho^-\to jj)\approx&~N_c\times\Gamma(\rho^-\to\ell^-\bar\nu_\ell)\approx N_fN_c\frac{g^2M_\rho}{48\pi g_\rho^2},\\
\Gamma(\rho^-\to W^-Z)\approx&~\Gamma(\rho^-\to W^-h)\approx\frac{a_\rho^4g_\rho^2M_\rho}{192\pi},
\end{split}\ee
where $N_f=3$, $N_c=3$, and the parameter $a_\rho=m_\rho/(g_\rho f)$. Similar results can be obtained in the $\rho^0$ decay. As illustrated in Fig.~\ref{fig:rho_decay}, this analytical approximation matches very well with the full numerical result, which is obtained by diagonalizing the mass matrices numerically. We have checked that the triple-boson channels, such as $\rho^+\to W^+ZZ$ and $W^+hh$, are around one order of magnitude smaller than the di-boson channels (by taking $M_\rho=4$ TeV and testing different $g_\rho$ values), thus we will not consider them in the following projected reach discussion.

\subsection{Projected reach}

Let's now turn to the expected reach on the parameter space $(M_\rho,g_\rho)$ in our model at the 10 TeV high energy muon collider with an integrated luminosity up to 10 ab$^{-1}$. According to the decay branching ratio discussed above, we will consider the di-boson and di-lepton/lepton-neutrino decay final states of $\rho$ to probe the large and small $g_\rho$ regions, respectively. For the $\rho V$ production channels, we consider the two processes
\be\label{rhoV_signal}
\mu^+\mu^-\to\rho^\pm(\to W^\pm Z)W^\mp,~\rho^0(\to W^+W^-)Z,
\ee
which result in the same $W^+W^-Z$ final state; and the leptonically decaying channel of the charged $\rho^\pm$ resonance,
\beq
\mu^+\mu^- \to \rho^\pm(\to e^\pm\nu_e) W^\mp.
\eeq
The $\rho^\pm\to\mu^\pm\nu_\mu$ decay channel has almost the same cross section as the electron channel, but it suffers from extra SM backgrounds such as the VBF production of the $W^\pm$ boson, thus we do not consider it here for simplicity. For the VBF production, we exploit the $W^+ W^-$ decay channel of the neutral resonance:
\be
\mu^+\mu^-\to \rho^0 (\to  W^+W^-)\nu_\mu\bar\nu_\mu.
\ee
while the cross section of charged ones is a factor of $\sim2$ smaller, as illustrated in Fig.~\ref{fig:rho_production}.

\begin{figure}
\centering
\subfigure{
\includegraphics[scale=0.33]{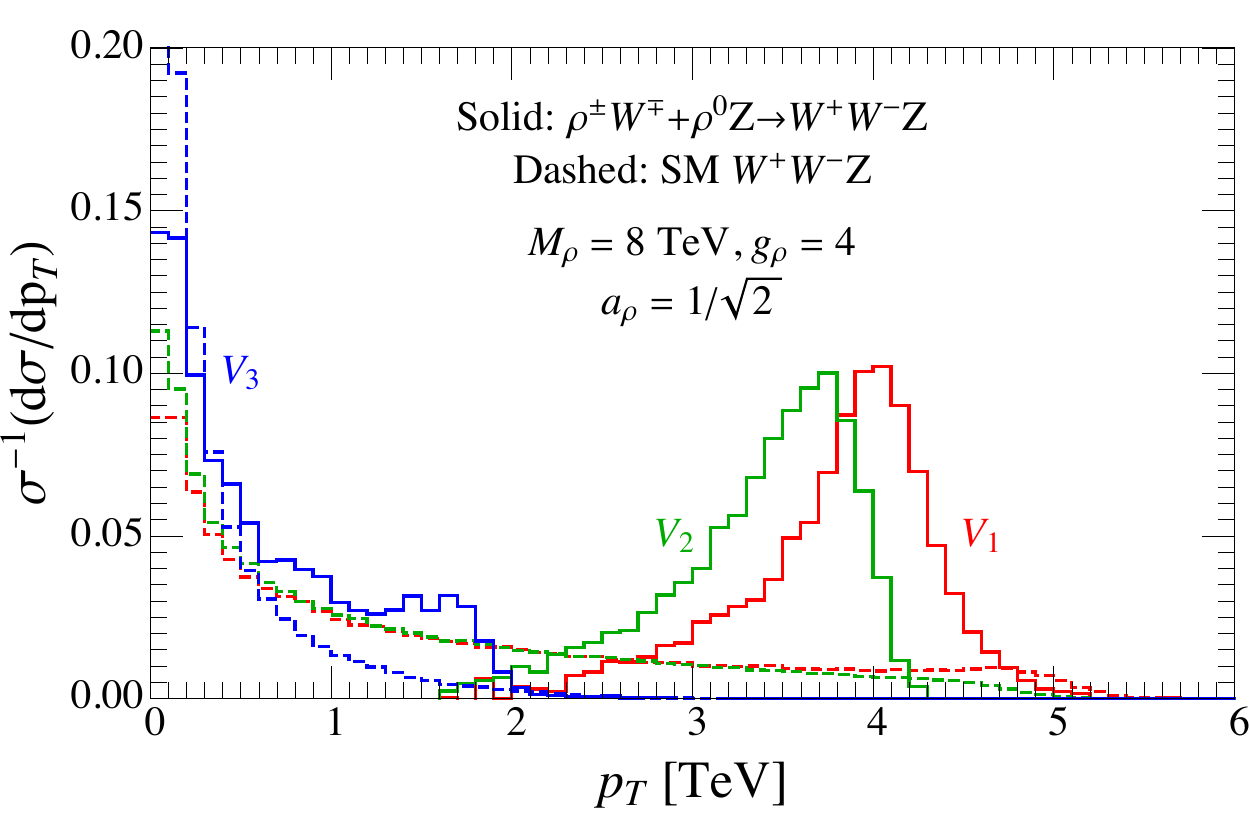}}
\subfigure{
\includegraphics[scale=0.33]{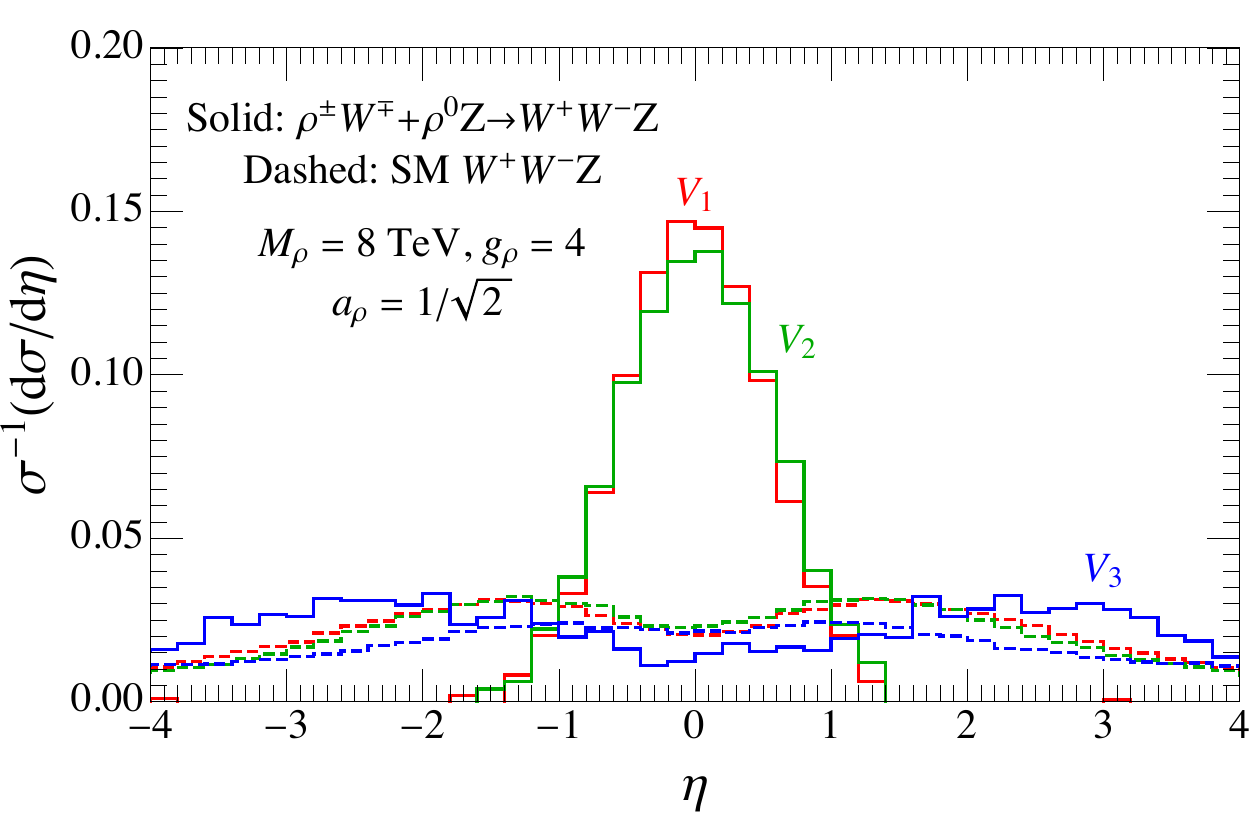}}
\caption{The $p_T$ and $\eta$ distributions of the signal and main background process of the $W^+W^-Z$ channel for $M_\rho=8$ TeV and $g_\rho=4$, before applying any selection cuts.}
\label{fig:distributions}
\end{figure}

Since we are mainly interested in the regime of $M_\rho \gtrsim $  TeV, the final state vector bosons from the $\rho$ decay are highly boosted so that they can be treated as fat-jets. We perform the event simulation at parton level with the {\tt MadGraph5aMC$@$NLO} event generator~\cite{Alwall:2014hca}. Although no decays of the $W^\pm$, $Z$ bosons are simulated, we consider the hadronically decaying channels of the $W^\pm$, $Z$ bosons by simply multiplying the decay branch ratios, BR$(W\rightarrow j j) \approx 67\%$ and BR$(Z\rightarrow j j) \approx 70\%$~\cite{ParticleDataGroup:2022pth}, and applying tagging efficiencies.  As a conservative estimate, we only make use of the hadronic decay channels. The leptonic decay channels of the $W^\pm$ and $Z$ bosons can be combined to further improve the sensitivity. In Fig.~\ref{fig:distributions}, we have plotted the event distributions for transverse momentum $p_T$ and the pseudorapidity $\eta$ of the three SM gauge bosons for the benchmark $M_\rho=8$ TeV and $g_\rho=4$. Here $V_{1,2,3}$ are ordered by their $p_T$. We can see that the leading and sub-leading $V$'s of the signal process are very central, and show clear Jacobi peaks at $p_T\sim M_\rho/2=4$ TeV. The third $V$ is softer, but still harder than the SM background.

We select the events with boosted jets with
\be\label{boosted_criterion}
p_T>500~{\rm GeV},\quad |\eta|<2.44,
\ee
separated by $\Delta R>1.0$, and adopt the tagging efficiencies listed in Appendix~\ref{app:tag}. The tagging and mis-tagging rates are taken from Ref.~\cite{CMS-DP-2023-065} from the simulations by the CMS collaboration using the Boosted Event Shape Tagger (BEST) at the 13 TeV LHC. In addition, we have applied  5\% energy smearing on the momenta of the jets. We require three boosted $V$-jets for the $\rho^\pm W^\mp/\rho^0Z\to W^+W^-Z$ channel, two boosted $V$-jets for the $\rho^0\nu_\mu\bar\nu_\mu\to W^+W^- \nu_\mu \bar\nu_\mu$  channel, and one boosted $V$-jet for the  $\rho^\pm W^\mp\to e^\pm\nu_e W^\mp$ channel. For the $e^\pm\nu_e W^\mp$ channel, we also require the lepton to be within \Eq{boosted_criterion}. We denote those selection requirements collectively as ``basic cut".

\begin{table}[h!]
\footnotesize\renewcommand\arraystretch{1.5}\centering
\begin{tabular}{|c|c|c|c|c|c|c|}\hline
\tabincell{c}{Cross section\\ $[{\rm fb}]$}  & \tabincell{c}{$\rho^\pm W^\mp/\rho^0Z$\\ $M_\rho=6$ TeV, $g_\rho=4$} & $W^+W^-Z$ & $ZZZ$  & $Vjj$ &  \tabincell{c}{Significance\\ at 100 fb$^{-1}$} \\ \hline
Before cuts & $8.28\times10^{-1}$ & 9.46 & $8.43\times10^{-2}$ & 4.07 & $-$ \\ \hline
Basic cuts & $5.49\times10^{-2}$ & $2.16\times10^{-1}$ & $9.81\times10^{-4}$ & $6.31\times10^{-3}$ & 1.12 \\ \hline
Mass shell cuts & $4.41\times10^{-2}$ & $5.93\times10^{-3}$ & $5.78\times10^{-5}$ & $3.89\times10^{-4}$ & 3.47 \\ \hline 
\end{tabular}
\caption{The cut flow of the signal and background of the $\rho^\pm W^\mp/\rho^0Z\to W^+W^-Z$ channel, where ``before cuts'' denotes the cross sections after the generator-level cuts but before selection cuts. The $Vjj$ process is calculated with $p_T^j>400$ GeV and $|\eta_j|<2.5$ to avoid IR divergence and improve event generation efficiency.}\label{tab:rhoV_pheno1}
\end{table}

To further suppress the background and enhance the signal significance, we try to reconstruct the resonance and select events with fat-jet invariant mass or recoil mass within the range of
\be\label{mjj_range}
M\in[M_\rho-2\Delta_\rho,M_\rho+2\Delta_\rho],
\ee
where $\Delta_\rho$ is $\Gamma_\rho$-dependent and defined as follows
\be\label{DeltaRho}
\Delta_\rho=\begin{dcases}
~0.05 M_\rho,&\Gamma_\rho<0.05\, M_\rho;\\
~\Gamma_\rho,&0.05\, M_\rho<\Gamma_\rho<1.5~{\rm TeV};\\
~1.5~{\rm TeV},&\Gamma_\rho>1.5~{\rm TeV},
\end{dcases}
\ee
and the factor 0.05 is consistent with our jet energy smearing rate of 5\%. For different channels, we apply different event selection cuts to implement the above requirement.
\begin{enumerate}
\item The $\rho^\pm W^\mp/\rho^0Z\to W^+W^-Z$ channel. We require the final state to have three boosted $V$-jets, at least one of the pair-combinations yields an invariant mass $M_{jj}$ within \Eq{mjj_range}. We further require the leading jet to have $p_T>M_\rho/2-1~{\rm TeV}$ and the sub-leading jet to have $p_T>M_\rho/2-0.5~{\rm TeV}$.

\item The $\rho^\pm W^\mp\to e^\pm\nu_e W^\mp$ channel. We require the final state to have exactly one electron with $p_T>{\rm max}\left\{0.5~{\rm TeV},~M_\rho/2-0.5~{\rm TeV}\right\}$ and $|\eta|<2.44$, and one boosted $V$-jet. The recoil mass $M_{\rm recoil}=\sqrt{(p_{\mu^+\mu^-}-p_V)^2}$ should be in the range set by \Eq{mjj_range}.

\item The $\rho^0\nu_\mu\bar\nu_\mu\to W^+W^- \nu_\mu \bar\nu_\mu$ VBF channel. In addition to the recoil mass cut in Eq.~(\ref{eq:xscut}), we require the final state to have two boosted $V$-jets, and veto any other flavor-tagged jets. The invariant mass of the two $V$-jets are required to be in the range of \Eq{mjj_range}; and the leading jet should have $p_T>M_\rho/2-1~{\rm TeV}$ and the sub-leading jet should have $p_T>M_\rho/2-0.5~{\rm TeV}$.
\end{enumerate}
We collectively denote these cuts as ``mass-shell" cuts.

\begin{table}\footnotesize\renewcommand\arraystretch{1.5}\centering
\begin{tabular}{|c|c|c|c|c|}\hline
\tabincell{c}{Cross section\\ $[{\rm fb}]$}  & \tabincell{c}{$\rho^\pm W^\mp$\\ $M_\rho=8$ TeV, $g_\rho=2$} & $e^\pm\nu_eW^\mp$ & $Ze^+e^-$ & \tabincell{c}{Significance\\ at 100 fb$^{-1}$} \\ \hline
Before cuts & $3.44\times10^{-1}$ & 8.92 & $2.23\times10^{-1}$ & $-$ \\ \hline
Basic cuts & $1.11\times10^{-1}$ & 1.36 & $2.71\times10^{-3}$ & 0.94  \\ \hline
Mass-shell cuts& $9.08\times10^{-2}$ & $4.16\times10^{-3}$ & 0 & 6.42 \\ \hline
\end{tabular}
\caption{Cut flow of the $\rho^\pm W^\mp\to e^\pm\nu_eW^\mp$ channel.}\label{tab:rhoV_pheno2}
\end{table}

The cross sections after the basic cuts and mass-shell cuts for different channels for the signal with some benchmark values of ($M_\rho,g_\rho$) and the leading backgrounds are listed in Table~\ref{tab:rhoV_pheno1} for the $W^+W^-Z$ channel, Table~\ref{tab:rhoV_pheno2} for the $e^\pm\nu_e W^\mp$ channel, and Table~\ref{tab:VBF_pheno} for the $W^+ W^- \nu_\mu \bar \nu_\mu$ VBF channel. We can see that after the cuts the irreducible backgrounds are always the dominant ones. In the $W^+W^-Z$ and $e^\pm\nu_e W^\mp$ channels, the VBF productions irreducible backgrounds are also considered. For example, the inclusion of VBF $VVV$ process (such as $\mu^+\mu^-\to W^+W^-Z\mu^+\mu^-/W^+W^-Z\nu_\mu\bar\nu_\mu$) will slightly reduce the reach for $M_\rho<4$ TeV. However, we have checked that they can be reduced by one order of magnitude by requiring the recoiled mass $\sqrt{p_{\mu^+\mu^-}-p_{VVV}}<1$ TeV, while the signal remains almost unchanged. Therefore, the VBF $VVV$ processes have a negligible impact on the projected reach. On the other hand, the VBF $e^\pm\nu_e W^\mp$ process can be greatly removed by the mass-shell cuts and hence is negligible. For the $W^+ W^- \nu_\mu \bar \nu_\mu$ VBF channel, the VBF production of $VV$ (such as $\mu^+\mu^-\to W^+W^-\nu_\mu\bar\nu_\mu/W^+Z\mu^-\bar\nu_\mu$) dominates the backgrounds, and the charged-current fusion component can be efficiently removed by vetoing the charged leptons within $p_T>50$ GeV and $|\eta|<2.44$. The signal significance $Z_0$ is obtained by the following formula~\cite{Cowan:2010js}:
\be
Z_0 =\sqrt{2}\sqrt{(S+B)\log\left(1+\frac{S}{B}\right)-S},
\ee
which reduces to $S/\sqrt{B}$ in the limit $S \ll B$.

\begin{table}[h!]\footnotesize\renewcommand\arraystretch{1.5}\centering
\begin{tabular}{|c|c|c|c|}\hline
\tabincell{c}{Cross section\\ $[{\rm fb}]$}  & \tabincell{c}{$\rho^0 \nu_\mu\bar\nu_\mu$\\ $M_\rho=4$ TeV, $g_\rho=8$} & $VV$ (VBF) & \tabincell{c}{Significance\\ at 100 fb$^{-1}$}   \\ \hline
Before cuts & $5.48\times10^{-1}$ & $1.15\times10^3$ & $-$ \\ \hline
Basic cuts & $6.89\times10^{-2}$ & $4.93\times10^{-1}$ & 0.96 \\ \hline
Mass-shell cuts & $4.39\times10^{-2}$ & $8.98\times10^{-2}$ & 1.36 \\ \hline
\end{tabular}
\caption{Cut flow of the $\rho^0\nu_\mu\bar\nu_\mu\to W^+W^-\nu_\mu\bar\nu_\mu$ channel.}\label{tab:VBF_pheno}
\end{table}

\begin{figure}[h!]
\centering
\includegraphics[scale=0.4]{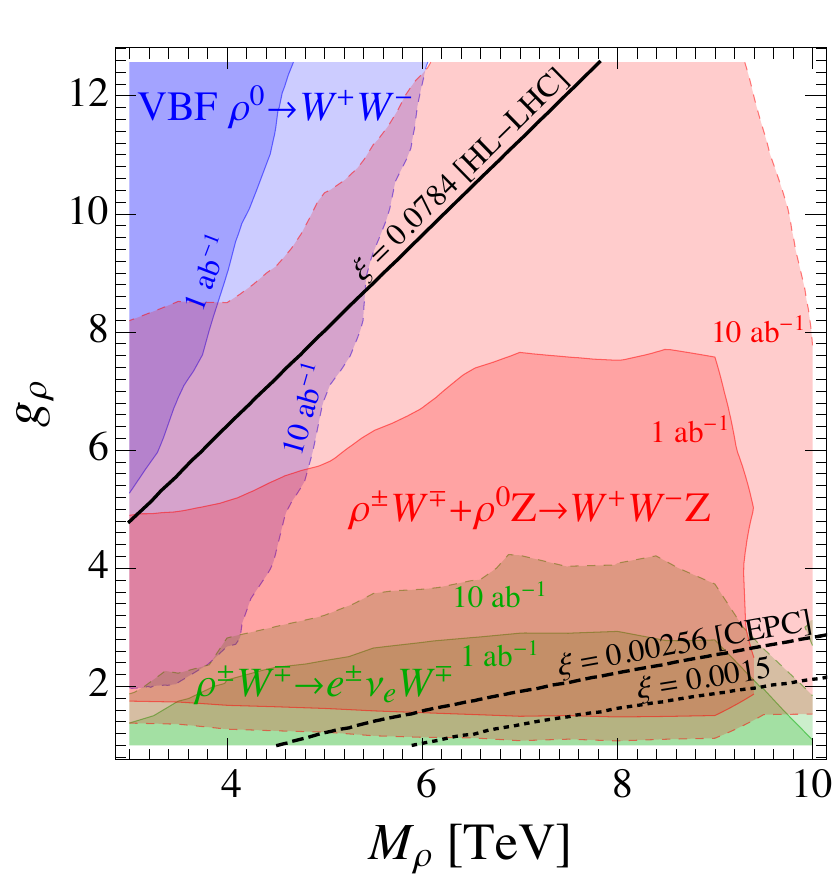}\qquad\qquad
\includegraphics[scale=0.425]{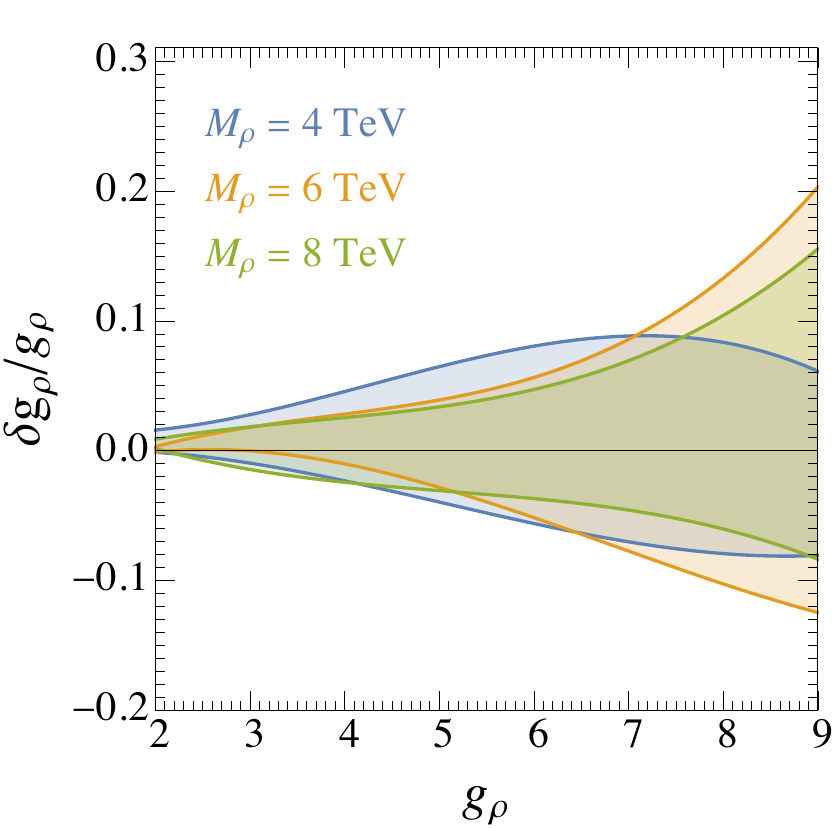}
\caption{Projections for the reach of composite $\rho$ resonance at a 10 TeV muon collider (left panel), and the measurement accuracy of the $g_\rho$ parameter, for different $M_\rho$ values (right panel). The limits on $\xi=v^2/f^2$ (lines in the left panel) are derived by assuming $a_{\rho}=m_\rho/(g_\rho f)=1/\sqrt{2}$.}
\label{fig:rho_projection}
\end{figure}

Based on the above analysis, we make the $5\sigma$ discovery projections on the $(M_\rho,g_\rho)$ parameter space at the 10 TeV muon collider with integrated luminosity of 1 ab$^{-1}$  and  10 ab$^{-1}$ in Fig.~\ref{fig:rho_projection}. 
With 1 ab$^{-1}$ integrated luminosity, the three channels studied above can already cover a large part of the parameter space, up to the kinematical limit for $m_{\rho}$ and a large region of the coupling $g_\rho$. We also notice that due to the overall large cross section, despite the suppression of $g^4/g_\rho^2$, $\rho V$ production with the $W^+W^-Z$ final state can cover the largest bulk of the parameter space and the uncovered space can be probed by VBF production with $W^+W^- \nu_\mu \bar \nu_\mu$ final state (large $g_\rho$, small $M_\rho$) and by the $\rho^\pm W^\mp$ with leptonic decay channel  $ e^\pm\nu_e W^\mp$ (very small $g_\rho \lesssim$ 1.5). The region covered by more than one color can be probed by combining different channels. For comparison, we have also shown the 95\% C.L. limit on the parameter $\xi = v^2/f^2 \approx a_\rho^2g_\rho^2 v^2/m_\rho^2$ from HL-LHC (0.0784)~\cite{CMS-NOTE-2012-006,ATL-PHYS-PUB-2018-054,Thamm:2015zwa}, CEPC (0.00256)~\cite{CEPCStudyGroup:2018ghi,An:2018dwb} (see similar projection at FCC-ee~\cite{FCC:2018evy,Thamm:2015zwa}) and from a 10 TeV high energy muon collider (0.0015)~\cite{Han:2020pif,Buttazzo:2020uzc}.

In the right panel of Fig.~\ref{fig:rho_projection}, we show the expected precision of the measurement of the coupling $g_\rho$ for $M_\rho = 4$, 6, and 8 TeV. As shown in the plot, we can achieve percent-level precision for $g_\rho \lesssim 5$ and 10\% precision for large $g_\rho \gtrsim 6$. The uncertainty of measuring $g_\rho$ is mainly determined by the $\rho^\pm W^\mp/\rho^0Z\to W^+W^-Z$ channel. At the same time, the VBF channels can also make important contributions. For $M_\rho=4$ TeV, large $g_\rho$ can be efficiently measured by the VBF process, thus $\delta g_\rho/g_\rho$ is constrained within $\leqslant10\%$ even for $g_\rho\gtrsim 8$; however, for $M_\rho=6$ and 8 TeV, the uncertainty for $g_\rho$ becomes larger since the VBF channel $\rho^0\nu_\mu\bar\nu_\mu\to W^+W^-\nu_\mu\bar\nu_\mu$ has a negligible signal significance. On the other hand, in the large $M_\rho$ region the leptonic decay channels $\rho^\pm W^\mp\to e^\pm\nu_e W^\mp$ provide a good probe for small $g_\rho$.

\section{The fermionic resonances (top partners)}
\label{sec:spin12}

In this section, we turn our discussion to the fermionic composite resonances. We will focus on the resonances that mainly couple to the top sector (the top partners), as these are expected to be the lightest ones under consideration of naturalness~\cite{Matsedonskyi:2012ym,Marzocca:2012zn}. We will only consider the quartet of $SO(4)$ and assume that the SM third generation quark doublet $q_L = (t_L, b_L)$ and the right-handed top quark $t_R$ belong to the elementary sector and are embedded in the \5 representation of $SO(5)$. We refer to the Appendix~\ref{app:model_fermion} for the detailed description of the effective Lagrangian under our consideration.

\subsection{Production and decay}

Let's start from the mass spectrum of the top partners. The top partners in $\4_{2/3} $ representation of $SO(4)\times U(1)_X$ can be decomposed to two  $SU(2)_L$ vector-like quark (VLQ) doublets, 
 \beq
 Q=\left(T,B\right)^T_{1/6},\quad Q_X=\left(X_{5/3},X_{2/3}\right)^T_{7/6}.
 \eeq
 where the SM hypercharges in the subscripts are determined by the combination of the third generator of $SU(2)_R$ and the $U(1)_X$ as $Y = T_R^3 + X$. Before the EW symmetry breaking (EWSB), there is a linear mixing between the left-handed components of the fermionic doublet, $Q_L$, and the SM left-handed third generation quark doublet $q_L = (t_L,b_L)_{1/6}$. As a result, the masses of the heavy VLQs $Q$ are given by
\beq
M_{T} = M_B =\sqrt{M_\Psi^2+y_L^2f^2},
\eeq
while the other doublet $Q_X$ have the masses
\beq
M_{X_{2/3}} = M_{X_{5/3}} = M_{\Psi}.
\eeq
After the EWSB, there will be additional $\mO(\sqrt{\xi})$ mixing between the right-handed top quark $t_R$ and the right-handed heavy charge-2/3 quarks $T_R$, $X_{2/3 R}$. After diagonalizing the mass matrix at at $\mO(\xi)$, the masses of the three charge-2/3 fermions can be written as 
\be\label{eq:topmass}\begin{split}
&M_t=\frac{y_Ly_Rf^2\sqrt{\xi}}{\sqrt{2}\sqrt{M_\Psi^2+y_L^2f^2}},\\
&M_{T}=\sqrt{f^2 y_L^2+M_\Psi^2}+\frac{M_\Psi^2 y_R^2f^2\xi}{4 \left(f^2
   y_L^2+M_\Psi^2\right)^{3/2}},\quad M_{X_{2/3}}=M_\Psi+\frac{y_R^2f^2\xi}{4 M_\Psi}.
\end{split}\ee
 Therefore, there are positive $\mO(\xi)$ corrections to the masses of $T$, $X_{2/3}$, but the mass formulae of $X_{5/3}$, $B$ remain the same as before the EWSB. In the limit of small $\xi$, we expect that the charge-5/3 resonance $X_{5/3}$ is the lightest particle in the quartet.

\begin{figure}[h!]
\centering
\subfigure{
\includegraphics[scale=0.85]{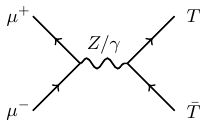}}\qquad
\subfigure{
\includegraphics[scale=0.85]{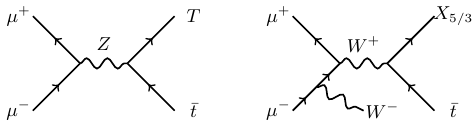}}\qquad
\subfigure{
\includegraphics[scale=0.85]{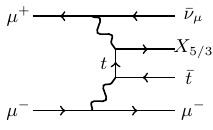}}
\caption{Feynman diagrams for the VLQ production Feynman. }\label{fig:VLQ_production_diagrams}
\end{figure}

\begin{figure}[h!]
\centering
\subfigure{
\includegraphics[scale=0.4]{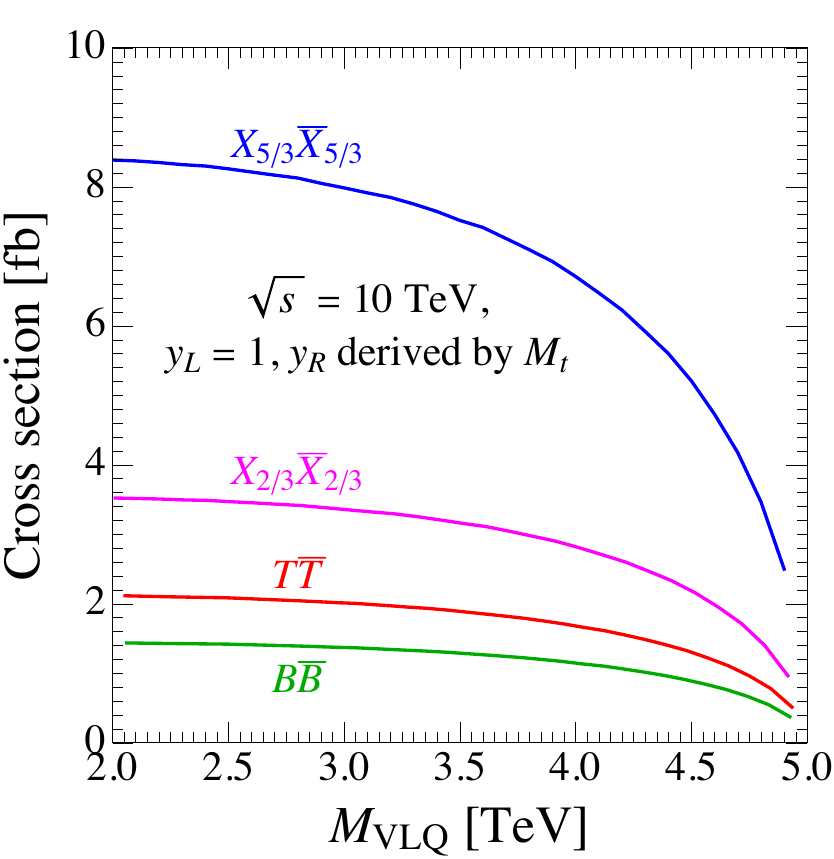}}\qquad\qquad
\subfigure{
\includegraphics[scale=0.4]{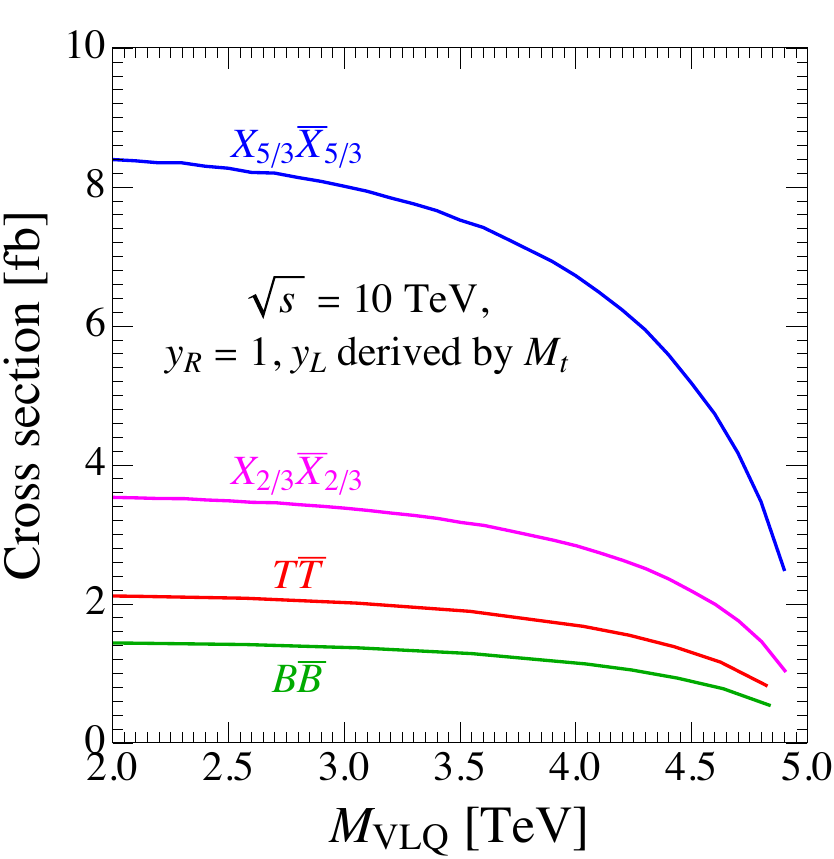}}
\subfigure{
\includegraphics[scale=0.4]{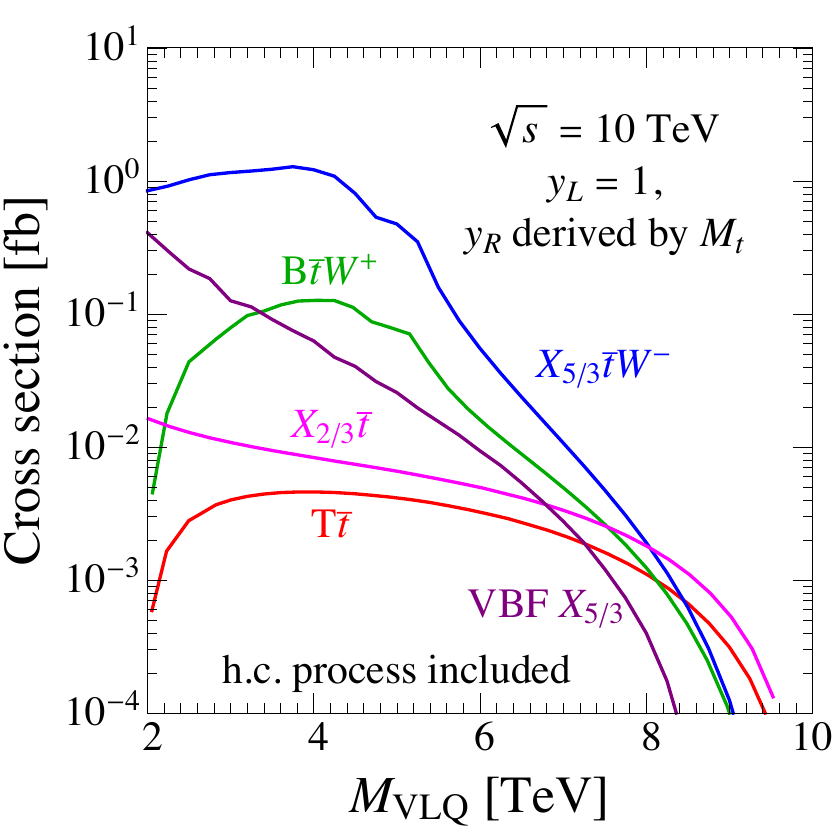}}\qquad\qquad
\subfigure{
\includegraphics[scale=0.4]{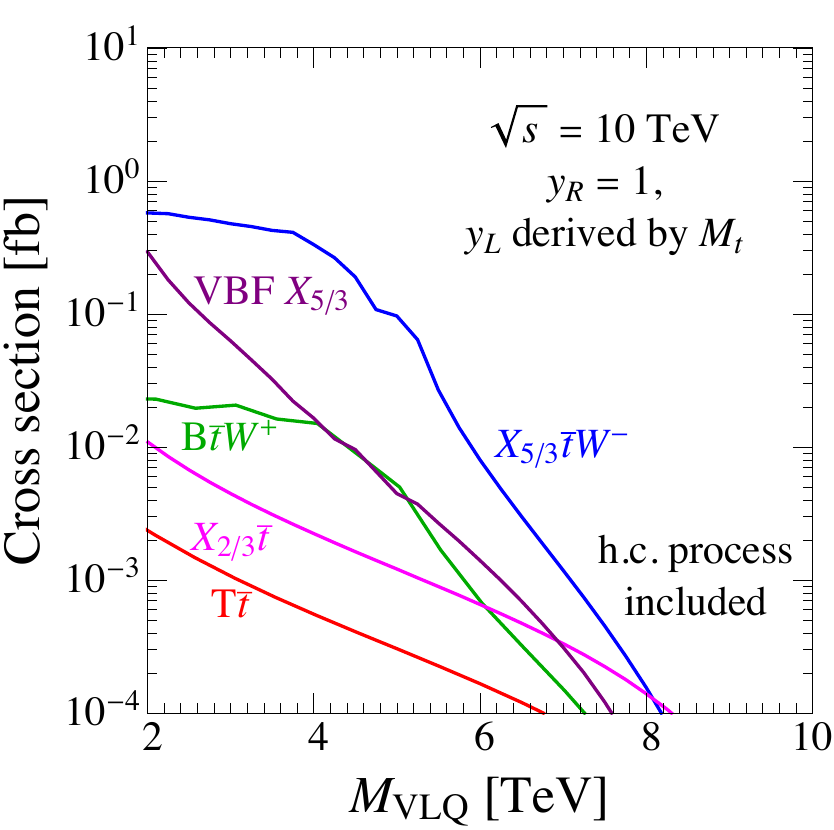}}
\caption{Top partner production rates as functions of their masses at a 10 TeV muon collider, with the pair (single) production channels shown in the top (bottom) row, respectively.   In the plots, we have fixed $f=2$ TeV. Note that the Lagrangian mass parameter $M_\Psi$ is different for different top partners. For all the single production modes, the charge-conjugate processes are also included.  When calculating the VBF cross sections, we require $M_{\mu^\pm\nu_\mu}>1$ TeV to remove the contribution from $tW$ fusion and $X_{5/3}$ pair production.}
\label{fig:VLQ_production}
\end{figure}

Because of their EW charges, all the VLQs can be produced in pairs by the $s$-channel $Z/\gamma$ exchange (see Fig.~\ref{fig:VLQ_production_diagrams}). In the limit of $M_{\Psi} \gg m_Z$, the rates are proportional to
\beq
\left(\frac{g^2}{2} T_{L,\rm VLQ}^3 + \frac{g^{\prime 2}}{2} Y_{\rm VLQ}\right)^2, \qquad Y^2_{\rm VLQ} g^{\prime 4},
\eeq
for the left-handed muon (right-handed anti-muon) initial states and the right-handed muon (left-handed anti-muon) initial states respectively. Here $T_{L,\rm VLQ}^3$, $Y_{\rm VLQ}$ represents the charge of $SU(2)_L$ isospin and the hypercharge for the heavy VLQs. Note that although the cross sections of the heavy quarks $T$, $B$ are dominated by the left-handed muon production, there are significant contributions from right-handed muon production for the $X_{5/3}$, $X_{2/3}$ because of their large hyper-charges. This explains the larger production cross sections for $X_{5/3}$, $X_{2/3}$  in comparison with those of $T$, $B$  in the upper panels of Fig.~\ref{fig:VLQ_production}. Note that for different top partners, the Lagrangian mass parameter $M_{\Psi}$ corresponding to the same physical mass is different. In the top-left panel, we have fixed $y_L=1$, and derived $y_R$ by the requirement of reproducing the top running quark mass $M_t=150$ GeV. Hence,
\beq\label{eq:topmassyR}
y_R \approx \frac{y_t}{y_L } \frac{\sqrt{M_\Psi^2 + y_L^2 f^2}}{ f},
\eeq
whose values is $\sim 1.2-2.3 $ for $M_\Psi \sim 2 - 5 $ TeV. In the top-right panel of Fig.~\ref{fig:VLQ_production}, we have fixed $y_R=1$ and  $y_L$ is fixed by the top quark mass requirement as
\beq\label{eq:topmassyL}
y_L \approx \frac{y_t}{\sqrt{y_R^2 - y_t^2} } \frac{M_\Psi}{ f} ,
\eeq
whose values is $\sim 1.8-4.2 $ for $M_\Psi \sim 2 - 5$ TeV. Here, the top Yukawa coupling is  $y_t = \sqrt{2} M_t /v$, and we have chosen $f = 2$ TeV. We find that \Eq{eq:topmassyR} and \Eq{eq:topmassyL} match the numerical results very well.

For the pair production, near the kinematical threshold $M_{\rm VLQ} \sim \sqrt{s}/2$, there is a two-body phase space suppression~\cite{Han:2005mu}, which depends on  the velocity of the heavy VLQ 
\beq
\beta_{\rm VLQ} = \sqrt{1 - \frac{4 M_{\rm VLQ}^2}{s}}.
\eeq
This explains the sharp falling of the cross sections when $M_{\Psi}$ approaches the kinematic limit of 5 TeV. The top partners can also be produced in pairs from VBF processes. Due to their smallness, we will not pursue it further.

In addition to the EW pair production, the VLQs can also be singly produced in association with one or two SM particles. As shown in Fig.~\ref{fig:VLQ_production_diagrams}, the charge-$2/3$ VLQs  can be produced together with one top/anti-top quark via Drell-Yan-like processes
\be
\mu^+\mu^-\to T \bar t,~X_{2/3}\bar t+\hc~.
\ee
For the charge-$5/3$ and $-1/3$ VLQs, an extra $W^\pm$ radiation is needed to conserve the electric charge. We have
\be
\mu^+\mu^-\to X_{5/3}\bar tW^-,~B\bar tW^++\hc~.
\ee
The lack of $B\bar b$ production is due to the $P_{LR}$ parity in the embedding of SM left-handed $q_L$ doublet~\cite{Agashe:2006at}. As shown in the bottom panels of Fig.~\ref{fig:VLQ_production} for the 10 TeV muon collider, although single production can potentially be used to get closer to the kinematic limit $M_{\rm VLQ}\sim\sqrt{s}$, the rates are rather small. This is due to the smallness of the coupling $ZTt$, $ZX_{2/3}t$, $X_{5/3} t W$, and $B t W$ (arising after the EWSB as $\mO( g \sqrt{\xi})$~\cite{Liu:2018hum}), and phase space suppression for the three-body $X_{5/3} t W$ and $B t W$ processes. Note that in our study, we work at the LO matrix element level. More accurate predictions require resumming the leading logs~\cite{Han:2020uid,Chen:2022msz,Garosi:2023bvq}. Similar to the pair production, in the bottom-left panel, we have fixed $y_L=1$ and derived $y_R\sim1.2-4.2$ by the requirement of reproducing the correct top quark mass; while in the bottom-right panel, we have fixed $y_R=1$ and derived $y_L\sim1.8-8.2$ in a similar way. In addition, for the $X_{5/3}$ and $B$ production channels, to remove the on-shell pair production contribution to the cross section, we require the invariant mass of the final state $t$ and $W$ to be outside of a 10\% window around the VLQ mass. 

The decays of VLQs can be understood by the Goldstone equivalence theorem with the interaction terms in Eq.~(\ref{eq:VLQ_expansion})~\cite{DeSimone:2012fs}. Among the two-body decaying channels of the VLQs,  the decays $X_{5/3}\to tW^+$ and $B\to tW^-$ have around 100\% branching ratios. For the charge-$2/3$ VLQs, when $M_\Psi, y_L f \gg y_R v$, we can apply the perturbative calculation as in \Eq{eq:topmass}, which implies
\be
\Br(T \to tZ)\approx\Br(T \to th)\approx\Br(X_{2/3}\to tZ)\approx\Br(X_{2/3}\to th)\approx50\%.
\ee

\subsection{Projected reach}

In this subsection, we report the projected reach on the top partners at a 10 TeV muon collider with 10 ab$^{-1}$ integrated luminosity. We will focus on the charge-$5/3$ resonance, $X_{5/3}$, as it is usually the lightest top partner and decay exclusively into $tW^+$ final state. When its mass $M_{X_{5/3}} = M_{\Psi}$ is smaller than half of the center-of-mass energy of the muon pair $\sqrt{s}/2$, we will consider the pair production which leads to $t\bar t W^+ W^-$ final state
\be\label{X53_pair}
\mu^+\mu^-\to X_{5/3}\bar X_{5/3}\to tW^+\bar tW^-.
\ee
For $M_\Psi > \sqrt{s}/2$, we will need the single production which also ends in the $t\bar t W^+ W^-$ final state
\beq
\mu^+\mu^-\to X_{5/3}\bar tW^-,~\bar X_{5/3}tW^+\to tW^+\bar tW^-.
\eeq
We will consider the fully hadronic decaying channel of the $t \bar t W^+ W^-$, which has a branching ratio of $\approx20\%$. Similar to our study of the spin-1 resonances, we perform a parton-level analysis by  MadGraph5 simulation and using the boosted jet tagging efficiencies given in Appendix~\ref{app:tag}. 

\begin{table}[h!]\footnotesize\renewcommand\arraystretch{1.5}\centering
\begin{tabular}{|c|c|c|c|c|c|c|}\hline
\tabincell{c}{Cross section\\ $[{\rm fb}]$}  & \tabincell{c}{$X_{5/3}\bar X_{5/3}$\\ $M_\Psi=4.95$ TeV} & $t\bar tVV$  & $VVjj$  &  $t\bar tjj$ & \tabincell{c}{Significance\\ at 100 fb$^{-1}$} \\ \hline
Before cuts & 1.32 & $1.94\times10^{-1}$ & 3.25 & 2.72 & $-$ \\ \hline
Basic cuts & $8.65\times10^{-2}$ & $2.33\times10^{-3}$ & $6.56\times10^{-4}$ & $4.23\times10^{-4}$ & 6.45 \\ \hline
Mass shell cuts & $3.66\times10^{-2}$ & $1.64\times10^{-5}$ & 0 & 0 & 7.01 \\ \hline
\end{tabular}
\caption{Cut flow of the $X_{5/3}\bar X_{5/3}\to tW^+\bar tW^-$ channel. The $VVjj$ and $t\bar tjj$ cross section are computed with $p_T^j>400$ GeV and $|\eta_j|<3$ to avoid IR divergence.}
\label{tab:pair_pheno}
\end{table}
\begin{table}[h!]\tiny\renewcommand\arraystretch{1.5}\centering
\resizebox{\textwidth}{11mm}{
\begin{tabular}{|c|c|c|c|c|c|c|c|c|}\hline
\tabincell{c}{Cross section\\ $[{\rm fb}]$}  & \tabincell{c}{$X_{5/3}\bar tW^-+{\rm h.c.}$\\ $M_\Psi=5.5$ TeV, $y_L=1$} & $t\bar tVV$  & $VVjj$ &  $t\bar tjj$ & $VVVV$  & $Vjj$ & $VVV$ & \tabincell{c}{Significance\\ at 100 fb$^{-1}$} \\ \hline
Before cuts & $2.39\times10^{-1}$ & $1.94\times10^{-1}$ & 3.25 & 2.72 & 1.88 & 4.07 & 9.55 & $-$ \\ \hline
Basic cuts & $1.68\times10^{-2}$ & $1.12\times10^{-2}$ & $1.66\times10^{-3}$ & $2.26\times10^{-2}$ & $3.64\times10^{-6}$ &  $1.12\times10^{-3}$ & $3.01\times10^{-5}$ & 0.82 \\ \hline
Mass shell cuts & $9.59\times10^{-3}$ & $2.18\times10^{-3}$ &  $7.95\times10^{-5}$ & $2.54\times10^{-3}$ & $1.55\times10^{-6}$ & $2.49\times10^{-4}$ & 0 & 1.09 \\ \hline
\end{tabular}}
\caption{Cut flow of the $X_{5/3} \bar tW^-+\hc\to tW^+\bar tW^-$ channel. The $VVjj$ and $t\bar tjj$ cross section are computed with $p_T^j>400$ GeV and $|\eta_j|<3$ to avoid IR divergence.}
\label{tab:single_pheno}
\end{table}

For basic selection cuts, we require two boosted top-tagged jets and two boosted $W$-tagged jet for the pair production channel, and two boosted top-tagged jets and at least one boosted $W$-tagged jet for the single production channel. The boosted jets are required to satisfy
\beq
p_T > 500 \,\text{GeV}, \qquad |\eta| < 2.44,
\eeq
and separated by an angular distance of $\Delta R > 1.0$. The cross sections for the signal and main backgrounds before any cuts and after the basic cuts are presented in Table~\ref{tab:pair_pheno} for pair production with $M_{\Psi} = 4.95 $ TeV, 
 and in Table~\ref{tab:single_pheno} for single production with $M_{\Psi} = 5.5$ TeV. To further improve the sensitivity, for the pair (single) production, we require two pairs (one pair) of top-$W$ boosted jets to have invariant mass within $[M_\Psi-2\Delta_\Psi,M_\Psi+2\Delta_\Psi]$, where $\Delta_\Psi$ is defined the same way as in~\Eq{DeltaRho} with $\rho\to\Psi$. The cross sections after this mass-shell cut are also listed in Table~\ref{tab:pair_pheno} and Table~\ref{tab:single_pheno}. In addition, we have also shown the expected discovery significance $Z_0$  for the integrated luminosity of 100 fb$^{-1}$ in the last column of the Table~\ref{tab:pair_pheno} and Table~\ref{tab:single_pheno}.  The results show that,  even for $M_\Psi=4.95$ TeV, which is very close to the pair production kinematical limit, the signal significance $Z_0$ could be as large as 7.0 for $\mL=100~{\rm fb}^{-1}$. Therefore, we expect the $M_\Psi<5$ TeV region can be well probed by the pair production of $X_{5/3}$.  

\begin{figure}[h!]
\centering
\subfigure{
\includegraphics[scale=0.4]{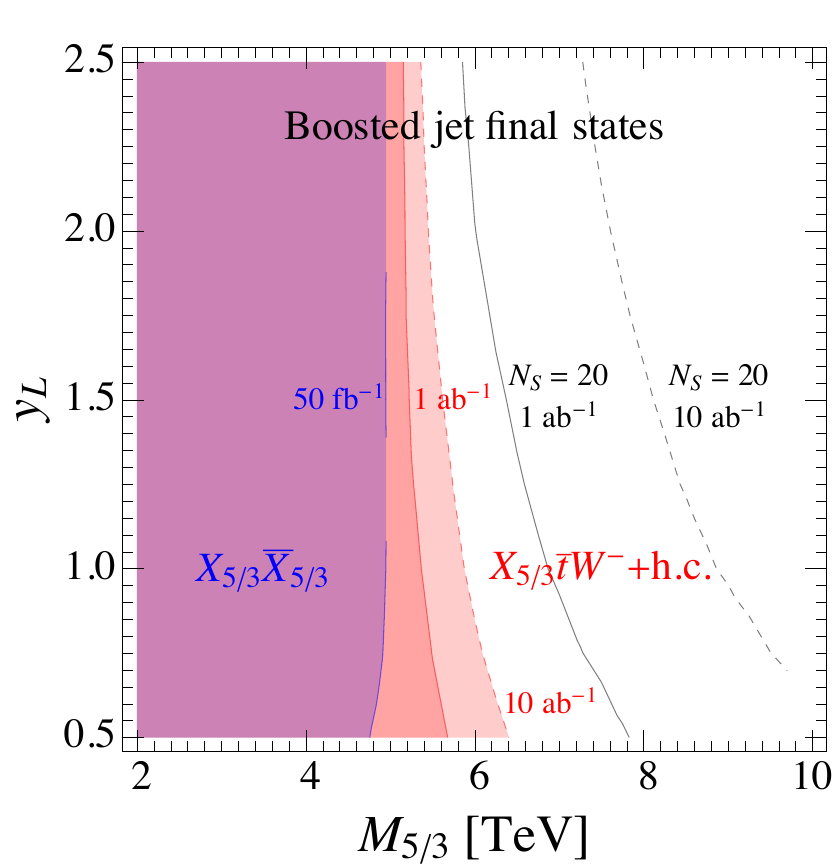}}
\caption{Projections for $5 \sigma$ discovery reach (colored contours) of composite top partner, $X_{5/3}$, at a 10 TeV muon collider. Both the pair production and single production are included. Through the pair production channel, we can reach the kinematical limit $M_{\Psi} < 5$ TeV. The single production channel can extend the reach to about 6 TeV. The black lines show the reach with 20 signal events. }
\label{fig:projection_VLQ}
\end{figure}

Based on the above analysis, we present the expected $5 \sigma$ discovery reach on the $M_{X_{5/3}} - y_L$ plane in Fig.~\ref{fig:projection_VLQ}, where the coupling $y_R$ is determined by the running top quark mass requirement $M_t = 150 $ GeV and $f$ is fixed to 2 TeV as usual.  For a broad range of couplings, $y_L$, we can cover almost all the parameter space for $M_{\Psi} < 5$ TeV, up to the kinematical limit of this channel. Beyond the kinematical reach of the pair production channel, we can use the single production processes to extend the reach to about $M_{\Psi} \sim 6$ TeV. The limitation on the potential of the single production channel mainly comes from the smallness of the production cross sections.  This is partially due to our choice of small $\xi$ as 0.015 ($f = 2$ TeV), because of the cross-section of the single production scales like $\xi$. We expect that the sensitivity can be enhanced if a larger value of $\xi$ is chosen, but we won't pursue it further here. In addition, we have only focused on a limited set of signals, and a combination of additional channels can certainly enhance the reach. In Fig.~\ref{fig:projection_VLQ}, we also show the potential reach with 20 signal events, which covers a much larger mass region. 

\begin{figure}[h!]
\centering
\includegraphics[scale=0.4]{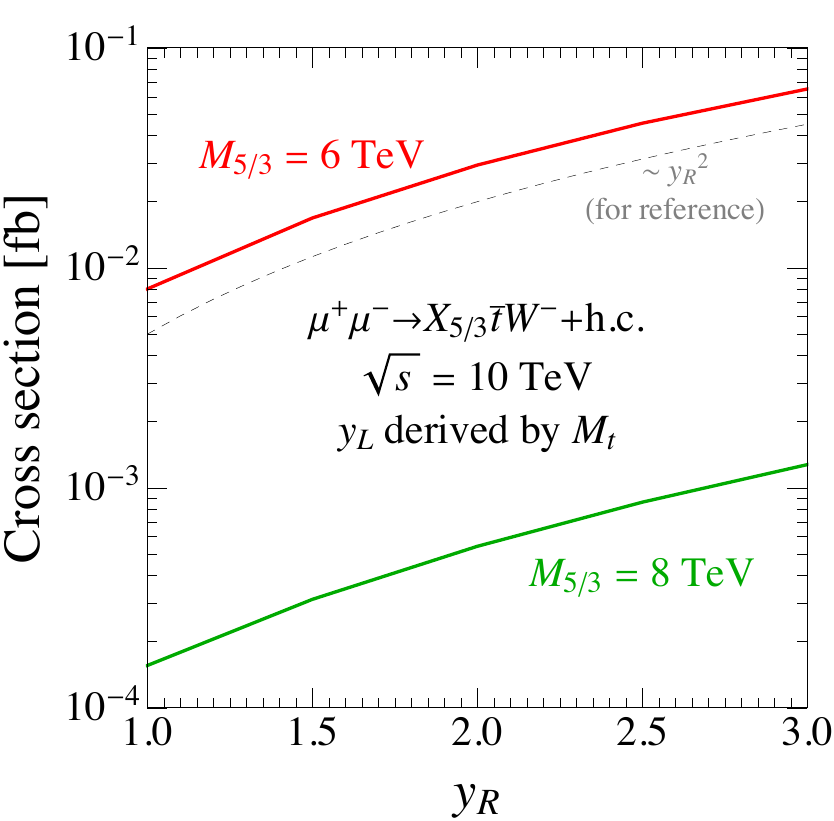}
\caption{The dependence of $X_{5/3}$ single production on $y_{R}$.}
\label{fig:X53_y_dependence}
\end{figure}

Before concluding this section, we would like to comment on the possibility of measuring the $y_{L,R}$ coupling once a discovery is made. In our model, the couplings $y_{L,R}$ are related to each other by the top quark mass, and we will focus on $y_R$. As shown explicitly in Fig.~\ref{fig:X53_y_dependence}, the cross sections of single production of $X_{5/3} tW$ scale as $y_R^2$ due to the mixing between $X_{2/3R}$ and $t_R$ at $\mO(\sqrt{\xi})$. This behavior should also hold for the VBF production of $X_{5/3} t$. In combination with other measurements, including the top partner mass, we can use the information of the production rate as a probe for coupling. We leave the prospect of measuring this coupling to future possible studies.

\section{Conclusion}

The physics potential at multi-TeV muon colliders is under active study. In this paper, we make a projection for the reach of the composite resonance searches at a 10 TeV muon collider with an integrated luminosity up to 10 ab$^{-1}$. We have focused on composite vector resonances in the $(\bold 3,\bold 1)$ of $SO(4)$, and fermionic resonances $(\bold 2, \bold 2)$ of $SO(4)$ in the MCHM, with the third-generation SM elementary quarks embedded in the \5 representation of $SO(5)$. The heavy resonances can be paired or singly produced in the  $\mu^+ \mu^-$ annihilation and VBF channels. After presenting the cross sections of different production channels, we make projections for the discovery reach on the parameter space by focusing on the $W^+W^-Z, e^\pm\nu_e W^\mp, W^+W^- \nu_\mu \bar\nu_\mu$ final states with hadronically decaying gauge bosons, resulting from $\mu^+\mu^-$ annihilation production of $\rho^\pm W^\mp/\rho^0Z$, $\rho^\pm W^\mp\to e^\pm\nu_e W^\mp$ and VBF production of  $\rho^0\nu_\mu\bar\nu_\mu$ respectively. For the fermionic resonances, we have considered the 
$ X_{5/3}\bar tW^-+\hc\to tW^+\bar tW^-$ channels with hadronically decaying top quarks. Our study shows that a 10 TeV muon collider with 10 ab$^{-1}$ luminosity can cover most of the kinematically allowed masses and a broad range of couplings of the vector resonances. It can also measure the new strong coupling $g_\rho$ to a few-percent level for small $g_\rho$ and tens of percent level for large $g_\rho$. For the fermionic resonances,  we can easily cover the parameter space below 5 TeV through the pair production channel. The single production can further extend the reach to 6 TeV top partners for a small  $\xi = 0.015$. 

Our work can be extended in several directions. Firstly, in this work, we considered vectorial resonances and fermionic resonances separately.  However, in a realistic model, both of these resonances should be present. As already pointed out in the literature, this can affect phenomenology significantly. For example, one kind of resonances can decay to the other, and vice versa.   It would be valuable to take this into account to arrive at a more complete picture. Secondly, we have studied several leading discovery channels in different sectors and it is certainly helpful to consider other production channels and decay modes. For example, by studying all the top partners, one can in principle infer the value of $y_L f$ from their mass measurement and by comparing their pair-production cross sections, we can confirm their EW quantum numbers. Lastly, in this paper, we have worked at the LO matrix element level to perform the simulation. At the same time, there has been a lot of progress toward developing the EW parton distributions at the high energy lepton collider. It is necessary to compare the results of fixed-order calculation to the results with resummed logarithms. We hope to address these issues in future work. 

\begin{acknowledgments}
We would like to thank Tao Han, Andrea Wulzer, and Keping Xie for useful discussions. We also thank Samantha Abbott, and Robin Erbacher for bringing Ref.~\cite{CMS-DP-2023-065} to our attention. The work of DL was supported in part by the U.S.~Department of Energy under grant No.~DE-SC0007914, in part by the PITT PACC, and was also supported by DOE Grant Number DE-SC-0009999. The work of L.T.W. was supported by DOE grant DE-SC-0013642. 
 \end{acknowledgments}

\appendix
\section{The minimal composite Higgs model}\label{app:model}

In this appendix, we quote the main formulae of the MCHM. More details can be found in Ref.~\cite{Panico:2015jxa} and the references therein, or in the appendix of Ref.~\cite{Liu:2018hum}.

\subsection{Symmetry breaking pattern and the Goldstone fields}

The global symmetry breaking of the strong dynamics sector is $SO(5)/SO(4)$, and the four components of the SM Higgs doublet are embedded into the coset space. The $SO(5)$ generators are
\be\label{app:SO(5)}\begin{split}
[T_L^a]_{IJ} =& -\frac{i}{2} \left[ \frac12 \epsilon^{abc}\left(\delta^{bI}\delta^{cJ} -\delta^{bJ}\delta^{cI} \right)+ \left(\delta^{aI}\delta^{4J} - \delta^{aJ} \delta^{4I}  \right)\right],\\
[T_R^a]_{IJ} =& -\frac{i}{2} \left[ \frac12 \epsilon^{abc}\left(\delta^{bI}\delta^{cJ} -\delta^{bJ}\delta^{cI} \right)- \left(\delta^{aI}\delta^{4J} - \delta^{aJ} \delta^{4I}  \right)\right],\\
[\hat T^i]_{IJ}=& - \frac{i}{\sqrt{2}}\left(\delta^{\hat{a} I} \delta^{5J} - \delta^{\hat{a}J} \delta^{5I}\right),
\end{split}\ee
where $I,J = 1,\,\cdots,\,5$. The normalization is $\tr[T^AT^B]=\delta_{AB}$. The unbroken $SO(4)\simeq SU(2)_L \times SU(2)_R$ generators can be expressed as
\be
\label{eq:44}
T^a_{L/R} = \begin{pmatrix}
t^a_{L/R} & 0 \\
0  & 0 
 \end{pmatrix},
\ee
where $a=1$, 2, 3, and $t^a_{L/R}$ are $4\times 4 $ matrices.

$\hat T^i$ denote the four broken generators of $SO(5)/SO(4)$ with $i=1,\,\cdots,\,4$, and the corresponding Goldstone degrees of freedom form a quartet of $SO(4)$, dubbed as $\vec{h}=(h_1,h_2,h_3,h_4)^T$, and the Goldstone matrix is
\be
U(\vec{h})=e^{i\frac{\sqrt{2}}{f}h_i\hat T^i}=\begin{pmatrix}\mathbb{I}_{4\times4}-\left(1-\cos\frac{h}{f}\right)\frac{\vec{h}(\vec{h}^T)}{h^2}&\frac{\vec{h}}{h}\sin\frac{h}{f}\\
-\frac{\vec{h}^T}{h}\sin\frac{h}{f}&\cos\frac{h}{f}\end{pmatrix}, 
\ee
where $h=\sqrt{|\vec{h}|^2}$. Under the transformation $\mG\in SO(5)$, the Goldstone matrix transforms as $U(\vec{h})\to\mG U(\vec{h})\mH^{-1}(\vec{h};\mG)$, where $\mH\in SO(4)$. Based on this non-linear realization of $SO(5)$, we can define the Maurer-Cartan form as
\be
U^\dagger\left(g_0W_\mu^a T_L^a+g'_0B_\mu T_R^3+i\partial_\mu\right)U=d_\mu^i\hat T^i+e_{L\mu}^aT_L^a+e_{R\mu}^aT_R^a\equiv d_\mu+e_\mu,
\ee
i.e. only a subgroup $SU(2)_L\times U(1)_Y\subset SO(4)$ is gauged. The $d$ and $e$ symbols transform as $d_\mu\to\mH d_\mu\mH^{-1}$ and $e_\mu\to\mH(e_\mu+i\partial_\mu)\mH^{-1}$, and they can be used to build up the CCWZ Lagrangian~\cite{Coleman:1969sm,Callan:1969sn} of the MCHM. For example, the Goldstone kinetic term is
\be\label{GK}
\frac{f^2}{4}d_\mu^id^{i\mu}=\frac{f^2}{2|H|^2}\sin^2\frac{\sqrt{2}|H|}{f}D_\mu H^\dagger D^\mu H+\frac{f^2}{8|H|^4}\left(\frac{2|H|^2}{f^2}-\sin^2\frac{\sqrt{2}|H|}{f}\right)(\partial_\mu|H|^2)^2,
\ee
with $H=(h_2+ih_1,h_4-ih_3)^T/\sqrt{2}$ being the Higgs doublet, $|H|=\sqrt{|H|^2}$, and $D_\mu$ is the normal $SU(2)_L\times U(1)_Y$ covariant derivative. \Eq{GK} can be expressed as $D_\mu H^\dagger D^\mu H+$ higher dimensional operators.

\subsection{Vector sector: the $\rho^{\pm,0}$ resonances}

We consider vector resonance $\rho_\mu$ in $(\3,\1)$ of $SO(4)\simeq SU(2)_L\times SU(2)_R$. The relevant Lagrangian is~\cite{Contino:2011np}
\be\label{eq:vector}
\mL_\rho=-\frac{1}{4}\rho_{\mu\nu}^a\rho^{a\mu\nu}+\frac{m_\rho^2}{2g_\rho^2}\left(g_\rho\rho_\mu^a-e_{L\mu}^a\right)^2,
\ee
where $g_\rho$ is the strong dynamics coupling, and the field strength tensor is
\be
\rho_{\mu\nu}^a=\partial_\mu\rho^a_\nu-\partial_\nu\rho^a_\mu+g_\rho\epsilon^{abc}\rho_\mu^b\rho_\nu^c.
\ee
The Lagrangian \Eq{eq:vector} can be expanded as
\be
\mL_\rho=-\frac{1}{4}\rho_{\mu\nu}^a\rho^{a\mu\nu}+\frac{a_\rho^2}{2}f^2\left(g_\rho\rho_\mu^a-g_0W_\mu^a+\frac{1}{f^2}H^\dagger\frac{\sigma^a}{2}i\Dfbd H\right)^2+\cdots
\ee
where $a_{\rho}=m_{\rho}/(g_{\rho}f)$ and ``$\cdots$'' denotes higher order terms in $|H|^2/f^2$.

After the EWSB, the mass term of \Eq{eq:vector} becomes
\be
\mL_\rho\supset\begin{pmatrix}W_\mu^-&\rho^-_{\mu}\end{pmatrix}M_\pm^2\begin{pmatrix} W^{+\mu}\\ \rho^{+\mu}\end{pmatrix}+\frac{1}{2}\begin{pmatrix} B_\mu& W_\mu^3& \rho^3_{\mu}\end{pmatrix}M_0^2\begin{pmatrix}B^\mu\\ W^{3\mu}\\ \rho^{3\mu}\end{pmatrix},
\ee
where $\rho^{\mp}=(\rho^1\pm i\rho^2)/\sqrt{2}$, and 
\be\label{Mpm}
M_\pm^2=\left(
\begin{array}{cc}
 \frac{1}{4}g_0^2f^2\left[a_{\rho}^2 \left(1+\sqrt{1-\xi }\right)^2+\xi
   \right] & -\frac{1}{2} a_{\rho}^2 g_{\rho}g_0 f^2  \left(\sqrt{1-\xi
   }+1\right) \\
 -\frac{1}{2} a_{\rho}^2 g_{\rho}g_0 f^2 \left(\sqrt{1-\xi }+1\right) &
   a_{\rho}^2g_{\rho}^2f^2 \\
\end{array}
\right),
\ee
and
\be\label{M0}\begin{split}
&M_0^2=\\&{\footnotesize \left(
\begin{array}{ccc}
 \frac{1}{4} g_0'^2f^2  \left[\xi +a_{\rho}^2 \left(1-\sqrt{1-\xi }\right)^2\right]
   & \frac{1}{4} (a_{\rho}^2-1)g_0 g'_0  f^2 \xi  & \frac{1}{2} a_{\rho}^2 
   g_{\rho}g'_0f^2\left(\sqrt{1-\xi }-1\right) \\
 \frac{1}{4} (a_{\rho}^2-1) g_0 g'_0 f^2 \xi  & \frac{1}{4} g_0^2 f^2
   \left[a_{\rho}^2 \left(1+\sqrt{1-\xi }\right)^2+\xi \right] & -\frac{1}{2} a_{\rho}^2
   g_{\rho}g_0 f^2  \left(\sqrt{1-\xi }+1\right) \\
 \frac{1}{2} a_{\rho}^2 g_{\rho}g'_0 f^2 \left(\sqrt{1-\xi }-1\right) &
   -\frac{1}{2} a_{\rho}^2 g_{\rho}g_0f^2 \left(\sqrt{1-\xi }+1\right) &
   a_{\rho}^2 g_{\rho}^2f^2  \\
\end{array}
\right)}.
\end{split}\ee
Here $\xi=v^2/f^2$, and $v\equiv f\sin(\ave{h}/f)$.

By diagonalizing the mass matrices in \Eq{Mpm} and \Eq{M0} we get the mass eigenstates of the vector bosons as well as their mixings, i.e.
\be
\begin{pmatrix}W_\mu^\pm\\ \rho_\mu^\pm\end{pmatrix}\xrightarrow[\rm eigenstates]{\rm To~mass}U_\pm\begin{pmatrix}W_\mu^\pm \\ \rho_\mu^\pm\end{pmatrix},\quad
\begin{pmatrix}B_\mu\\ W_\mu^3\\ \rho_\mu^3\end{pmatrix}\xrightarrow[\rm eigenstates]{\rm To~mass}U_0\begin{pmatrix}A_\mu\\ Z_\mu\\ \rho_\mu^3\end{pmatrix},
\ee
such that $U_\pm^\dagger M_\pm^2U_\pm={\rm diag}\{M_W^2,M_{\rho^\pm}^2\}$ and $U_0^\dagger M_0^2U_0={\rm diag}\{0,M_Z^2,M_{\rho^0}^2\}$. The $U_\pm$ and $U_0$ matrices can be analytically solved. Defining
\be\label{gge}
g=\frac{g_0g_\rho}{\sqrt{g_\rho^2+g_0^2}},\quad g'=g'_0,\quad e=\frac{gg'}{\sqrt{g^2+g'^2}},
\ee
the mass eigenvalues of the spin-1 resonances are
\be
M_{\rho^\pm}^2\approx M_{\rho^0}^2=\frac{g_{\rho}^2}{g_{\rho}^2-g^2}m_{\rho}^2-\frac{g^2\xi}{4}\left(\frac{2m_{\rho}^2-g^2f^2}{g_{\rho}^2-g^2}\right)+\mO(\xi^2),
\ee
and the $\rho$-$W$ mixing angle is $\sin\theta\approx g_0/g_\rho\approx g/g_\rho$.

The EW observables are affected by the mixing between SM sector and the strong dynamics sector. For the SM gauge bosons, the masses are
\be\label{MV}\begin{split}
M_W^2=&~\frac{g^2}{4}v^2\left[1+\frac{g^2\xi}{2g_\rho^2}\left(1-\frac{g^2}{2a_\rho^2g_\rho^2}\right)+\mO(\xi^2)\right],\\
M_Z^2=&~\frac{g^2+g'^2}{4}v^2\left[1+\frac{g^2\xi}{2g_\rho^2}\left(1-\frac{g^2}{2a_\rho^2g_\rho^2}\right)+\mO(\xi^2)\right],
\end{split}\ee
and the photon is massless due to the residual $U(1)_{\rm em}$ invariance. The weak coupling is
\be
g_W=g
\left[1+\frac{g^2 \xi}{4 g_\rho^2}\left(1-\frac{g^2}{a_\rho^2g_\rho^2}\right)+\mO(\xi^2)\right],
\ee
and the Fermi constant is
\be\label{GF}
G_F=\frac{1}{\sqrt{2}v^2}\left(1-\frac{g^4 \xi }{4 a_\rho^2 g_\rho^4}+\mO(\xi^2)\right).
\ee
Using \Eq{gge}, \Eq{MV} and \Eq{GF}, we can change the input parameters as
\be
\{g_\rho,f,a_\rho,g_0,g_0',v\} \to \{g_\rho,f,a_\rho,\alpha_{\rm EW},M_Z,G_F\},
\ee
where the last three are fixed by the experiment
\be
\alpha_{\rm EW}=\frac{1}{127.9},\quad M_Z=91.188~{\rm GeV},\quad G_F=1.16637\times10^{-5}~{\rm GeV}^{-2}.
\ee

\subsection{Fermion sector: the charge-$5/3$, $2/3$ and $-1/3$ top partners}

\label{app:model_fermion}
The global $SO(5)$ symmetry is extended to $SO(5)\times U(1)_X$ to give the correct hypercharge to the fermions, and the gauged subgroup is $SU(2)_L\times U(1)_Y$ that $Y=T_R^3+X$. The relevant Lagrangian is
\be\label{L_fermion}
\mL_\Psi=\bar\Psi\left(i\slashed{\nabla}+\frac{2}{3}g'_0\slashed{B}\right)\Psi-M_\4\bar\Psi\Psi+y_L f \bar{q}_L^{\5I}U_{Ij}\Psi^j+y_R f \bar{t}_R^{\5I}U_{Ij}\Psi^j +\hc,
\ee
where the fermion resonances
\be
\Psi = \frac{1}{\sqrt{2}}\begin{pmatrix}
i B - i X_{5/3}, &
 B + X_{5/3}, &
i T + i X_{2/3}, &
 -T + X_{2/3} 
\end{pmatrix}^T,
\ee
is the $\4_{2/3}$ multiplet under $SO(4)\times U(1)_X$, $\nabla_\mu=\partial_\mu-ie_{L\mu}^at_L^a-ie_{R\mu}^at_R^a$ is the $SO(5)/SO(4)$ covariant derivative, and
\be
q_L^\5=\frac{1}{\sqrt{2}}\begin{pmatrix}ib_L,&b_L,&it_L,&-t_L,&0\end{pmatrix}^T,\quad t_R^\5=\begin{pmatrix}0,&0,&0,&0,&t_R\end{pmatrix},
\ee
are the $\5_{2/3}$ embeddings of the elementary quark doublet $q_L=(t_L,b_L)^T$ and singlet $t_R$ in $SO(5)\times U(1)_X$. The $y_{L,R}$ Yukawa terms in \Eq{L_fermion} are called ``partial compositeness'', which are crucial in generating SM top quark mass and the Higgs potential. The fermion resonances can couple to the vector resonances via
\be
\mL_{\rho\Psi}=c_1\bar\Psi\gamma^\mu t_L^a\Psi\left(g_\rho\rho_\mu^a-e_\mu^a\right),
\ee
where $c_1$ is an $\mO(1)$ number.

Under the $SO(4)\times U(1)_X\to SU(2)_L\times U(1)_Y$ decomposition, $\4_{2/3}\to\2_{7/6}\oplus\2_{1/6}$, thus the $\Psi$ quartet can be decomposed into two SM doublets, i.e.
\be
Q_X=\begin{pmatrix}X_{5/3}\\ X_{2/3}\end{pmatrix},\quad Q=\begin{pmatrix} T \\ B\end{pmatrix};
\ee
and correspondingly, the Lagrangian can be expanded as
\bea\label{eq:VLQ_expansion}
\mL_\Psi&=&\bar Q(i\slashed{D}-M_\4)Q+\bar Q_X(i\slashed{D}-M_\4)Q_X-\frac{1}{f^2}\left(\bar Q_X\gamma^\mu\frac{\sigma^a}{2}Q_X+\bar Q\gamma^\mu\frac{\sigma^a}{2}Q\right)H^\dagger \frac{\sigma^a}{2}i\Dfbd H\nn\\
&&-\frac{1}{4f^2}\left(\bar Q_X\gamma^\mu Q_X-\bar Q\gamma^\mu Q\right)H^\dagger i\Dfbd H+\left(\frac{1}{4f^2}\bar Q\gamma^\mu Q_XH^\dagger i\Dfbd\tilde H+\hc\right)\\
&&+y_Lf\left[\bar q_LQ_R+\frac{1}{2f^2}\bar q_L\tilde H\left(H^\dagger Q_{XR}-\tilde H^\dagger Q_R\right)\right]+y_R\left(\bar Q_L\tilde H-\bar Q_{XL}H\right)t_R+\hc+\cdots\nn
\eea
and
\be
\mL_{\rho\Psi}=c_1\left(\bar Q \gamma^\mu\frac{\sigma^a}{2}Q+\bar Q_X\gamma^\mu\frac{\sigma^a}{2}Q_X\right)\left(g_\rho\rho_\mu^a-g_0W_\mu^a+\frac{1}{f^2}H^\dagger\frac{\sigma^a}{2}i\Dfbd H\right)+\cdots
\ee
where ``$\cdots$'' denotes the higher order terms in $|H|^2/f^2$.

After the EWSB, \Eq{L_fermion} yields the fermion mass term
\be\begin{split}
\mL_\Psi\supset&~-\begin{pmatrix}\bar{b}&\bar{B}\end{pmatrix}_L
M_{-1/3}\begin{pmatrix}b\\ B\end{pmatrix}_R-\begin{pmatrix}\bar{t}&\bar{T}&\bar X_{2/3}\end{pmatrix}_LM_{2/3}\begin{pmatrix}t\\ T\\ X_{2/3}\end{pmatrix}_R+\text{h.c.}\\
&~-M_\4\bar X_{5/3}X_{5/3},
\end{split}\ee
where the charge-$5/3$ resonance has a mass $M_\4$, and the charge-$2/3$ and $-1/3$ fermions mix with the following mass matrices
\be\label{mass_matrix_fermion}
M_{-1/3}=\begin{pmatrix}0&-y_Lf\\0&M_\4\end{pmatrix},\quad
M_{2/3}=\begin{pmatrix}0&-\frac{y_Lf}{2}\left(1+\sqrt{1-\xi}\right)&-\frac{y_Lf}{2}\left(1-\sqrt{1-\xi}\right)\\
-\frac{y_Rf}{\sqrt2}\sqrt\xi&M_\4&0\\
\frac{y_Rf}{\sqrt2}\sqrt\xi&0&M_\4\end{pmatrix}.
\ee
For the charge-$-1/3$ fermions, the mass matrix can be analytically diagonalized by
\be
\begin{pmatrix}b\\ B\end{pmatrix}_L\xrightarrow[\rm eigenstates]{\rm To~mass}
U_b\begin{pmatrix}b \\ B\end{pmatrix}_L,\quad U_b=\begin{pmatrix}\frac{M_\4}{\sqrt{M_\4^2+y_L^2f^2}}&\frac{-y_Lf}{\sqrt{M_\4^2+y_L^2f^2}}\\
\frac{y_Lf}{\sqrt{M_\4^2+y_L^2f^2}}&\frac{M_\4}{\sqrt{M_\4^2+y_L^2f^2}}\end{pmatrix},
\ee
yielding a massless $b$ quark (because we haven't include a $b_R$ in \Eq{L_fermion}) and a vector-like quark with mass $M_B=\sqrt{M_\4^2+y_L^2f^2}$. The mass eigenstates for the charge-2/3 fermions can be obtained via the singularity value decomposition,
\be
\begin{pmatrix}t \\ T \\ X_{2/3}\end{pmatrix}_L \rightarrow
U_t\begin{pmatrix}t \\ T \\ X_{2/3}\end{pmatrix}_L,\quad
\begin{pmatrix}t\\ T \\ X_{2/3}\end{pmatrix}_R=
V_t\begin{pmatrix}t \\ T \\ X_{2/3}\end{pmatrix}_R,
\ee
with $U_t^\dagger M_{2/3}V_t={\rm diag}\{M_t,M_{T},M_{X_{2/3}}\}$, where $M_{T}>M_{X_{2/3}}$.

\section{Tagging and mistagging rates}\label{app:tag}

\begin{table}\footnotesize\renewcommand\arraystretch{1.5}\centering
\begin{tabular}{|c|c|c|c|c|c|c|}\hline
\diagbox{Truth}{Tagged} & $W$ & $Z$ & Higgs & Top & Bottom & QCD \\ \hline
$W$ & 0.73 & 0.17 & 0.01 & 0.03 & 0.02 & 0.04 \\ \hline
$Z$ & 0.22 & 0.64 & 0.05 & 0.03 & 0.03 & 0.04 \\ \hline
Higgs & 0.02 & 0.08 & 0.75 & 0.05 & 0.08 & 0.03 \\ \hline
Top & 0.04 & 0.03 & 0.04 & 0.81 & 0.04 & 0.04 \\ \hline
Bottom & 0.02 & 0.02 & 0.04 & 0.06 & 0.68 & 0.17 \\ \hline
QCD & 0.05 & 0.03 & 0.02 & 0.05 & 0.19 & 0.66 \\ \hline
\end{tabular}
\caption{Tagging and mistagging efficiencies of the boosted jets Taken from Ref.~\cite{CMS-DP-2023-065}.}\label{tab:tagging}
\end{table}

The tagging and mistaggin rates are taken from Ref.~\cite{CMS-DP-2023-065} from the simulations by the CMS collaboration using the Boosted Event Shape Tagger (BEST) at 13 TeV LHC, and listed in Table~\ref{tab:tagging}. Each horizontal row is normalized such that the efficiencies in the horizontal rows will sum to 1.

\bibliographystyle{JHEP-2-2.bst}
\bibliography{references}

\end{document}